\documentclass{jpp}
\usepackage{graphicx}
\usepackage{epstopdf, epsfig}
\usepackage{hyperref}
\usepackage{amsmath,amsfonts,amssymb}
\usepackage{subfigure}
\usepackage{relsize}
\usepackage{bm}
\usepackage{booktabs}
\usepackage{subcaption}

\usepackage{float}

\usepackage{fontawesome}

\graphicspath{{sources/figs}}

\usepackage{xcolor,soul}
\usepackage{hyperref}
\hypersetup{
  colorlinks   = true,
  urlcolor     = blue,
  linkcolor    = blue,
  citecolor   = blue
}

\newcommand{\EQ}{\begin{equation}}
\newcommand{\EN}{\end{equation}}
\newcommand{\EQA}{\begin{eqnarray}}
\newcommand{\ENA}{\end{eqnarray}}

\newcommand{\Eq}[1]{(\ref{#1})}
\newcommand{\Eqs}[2]{(\ref{#1}) and~(\ref{#2})}

\newcommand{\Sec}[1]{section~\ref{#1}}

\newcommand{\Fig}[1]{figure~\ref{#1}}
\newcommand{\FFig}[1]{Figure~\ref{#1}}

\newcommand{\TTab}[1]{Table~\ref{#1}}

\newcommand{\bra}[1]{\langle #1\rangle}

{}
{}

{}
{}

{}
{}
{}
{}
{}
{}
{}
{}
{}
{}
{}
{}
{}
{}

{}

{}

{}
{}
{}

\newcommand{\yyy}{\hat{\mbox{\boldmath $y$}} {}}

\newcommand{\BB}{\bm{B}}

\newcommand{\JJ}{\bm{J}}

\newcommand{\AAA}{\bm{A}}

\newcommand{\UU}{\bm{U}}

\newcommand{\nab}{{\bm{\nabla}}}

\newcommand{\SSSS}{\mbox{\boldmath ${\sf S}$} {}}

\newcommand{\DD}{{\rm D} {}}

\def\cs{c_{\rm s}}

\shorttitle{Prandtl number dependence of the reconnection rate}
\shortauthor{Vinay Kumar and Axel Brandenburg}

\title{Magnetic Prandtl number dependence of plasmoid-mediated reconnection}

\author{Vinay Kumar\aff{1,2}\corresp{\email{vinay.kumar@icts.res.in}, {the21vk@gmail.com}}
and Axel Brandenburg\aff{2,3,4,5}}

\affiliation{
\aff{1}International Centre for Theoretical Sciences, Tata Institute of Fundamental Research, Bangalore 560089, India
\aff{2}Nordita, KTH Royal Institute of Technology and Stockholm University,\\
Hannes Alfv\'ens v\"ag 12, SE-10691 Stockholm, Sweden
\aff{3}The Oskar Klein Centre, Department of Astronomy,
Stockholm University, AlbaNova, SE-10691 Stockholm, Sweden
\aff{4}McWilliams Center for Cosmology \& Department of Physics,
Carnegie Mellon University, Pittsburgh, PA 15213, USA
\aff{5}School of Natural Sciences and Medicine, Ilia State University,\\
3-5 Cholokashvili Avenue, 0194 Tbilisi, Georgia
}

\date{\today}

\begin{document}

\maketitle

\begin{abstract}
We investigate the dependence of the plasmoid-mediated magnetic reconnection rate on the magnetic Prandtl number using two-dimensional magnetohydrodynamic simulations of two coalescing magnetic islands. For Lundquist numbers below the onset of the plasmoid instability, the reconnection rate follows the expected Sweet–Parker scaling and decreases with increasing magnetic Prandtl number. However, once the current sheet becomes plasmoid unstable, the dependence on the magnetic Prandtl number weakens considerably. In the fully plasmoid-mediated regime, we find reconnection rates that remain nearly independent of the magnetic Prandtl number over the explored parameter range. 
We show that the largest reconnection rates are associated with strongly non-linear phases involving plasmoid interactions and mergers. 
We further compare our results with simulations of the boundary-driven Taylor problem, where previous studies reported a stronger magnetic Prandtl number dependence, and provide a possible explanation for the differing scalings obtained in the two setups. These results may have implications for reconnection-mediated decay in magnetically dominated turbulence and related astrophysical systems.

\end{abstract}

\keywords{astrophysical plasmas, plasma simulation, plasma non-linear phenomena}

\section{Introduction}

Magnetic reconnection is a fundamental mechanism for energy 
dissipation in plasmas. 
It enables the rapid conversion of magnetic energy into kinetic 
and thermal energy, and non-thermal particle populations through the 
reconfiguration of magnetic field topology \citep{Zweibel2009}. 
As such, it plays a central role across laboratory, space, and 
astrophysical environments 
\citep{Yamada1997, Paschmann2013,Priest2002}. 
Reconnection is widely invoked to explain a range of explosive 
phenomena, from solar and stellar flares and coronal mass 
ejections to magnetospheric substorms and high-energy particle 
acceleration \citep{Shibata2011,Angelopoulos2008,Sironi2025}.
It is important in large-scale astrophysical systems such as galaxy 
clusters, where reconnection has been proposed as a source of 
diffuse radio emission \citep{Ghosh2025}.

Magnetic reconnection occurs in localised regions with high
current density, where the magnetic field undergoes sharp
reversals. 
In the classical Sweet–Parker model, this process is described 
as occurring within an elongated current sheet of length $L$ and
thickness $\delta$, where 
oppositely directed magnetic fields with strength $B$ 
are brought together and reconnect due to finite, non-zero 
resistivity $\eta$. 
The resulting reconnection rate is constrained by the {aspect ratio} $\left(\sim \delta/L\right)$ 
of the diffusion region and scales as $S^{-1/2}$,
where {$S = BL/\eta\sqrt\rho = V_AL/\eta$} is the Lundquist number,
written here in units where the vacuum permeability is unity,
and $\rho$ is the density of the plasma \citep{Sweet1958, Parker1957}.

Typical astrophysical systems have very large Lundquist numbers, 
owing to their large spatial scales and extremely small 
resistivities. 
In this regime, the Sweet–Parker model predicts exceedingly {high-aspect-ratio}
current sheets and correspondingly slow reconnection rates,
rendering it incapable of explaining the rapid energy release
observed in many astrophysical phenomena.

This limitation has led to the exploration of several mechanisms for fast reconnection. One such mechanism is the \emph{plasmoid instability}: sufficiently long and thin current sheets become unstable and fragment into multiple magnetic islands and secondary current sheets, forming a hierarchical structure. This fragmentation enables multiple simultaneous reconnection sites and substantially enhances the overall reconnection rate, making it only weakly dependent on the Lundquist number \citep{Loureiro2007,Bhattacharjee2009,Huang2010,Uzdensky2010,Loureiro2012,Comisso2016}.
A similar result was obtained by \cite{GN96} in the context of flux-tube braiding at low plasma beta,
where the formation of a hierarchy of smaller current sheets causes the resulting heating rate to be independent of resistivity.
{A complementary perspective was proposed by \citet{Pucci2014}, who argued that at sufficiently large Lundquist numbers Sweet-Parker current sheets are super-Alfv\'enically unstable and therefore cannot form, with the tearing instability disrupting the current sheet before the classical Sweet-Parker state is established.}

Fast reconnection can also arise in turbulent environments, where stochastic field-line wandering broadens the effective outflow region and enhances magnetic flux transport across the reconnection layer \citep{Lazarian1999, Kowal2009,Kowal2012,Vicentin2025}. In addition, kinetic-scale processes may further modify the structure of the diffusion region, enabling fast reconnection in weakly collisional or collisionless plasmas
\citep{Bhat2018,Shay1999,Bessho2005,Drake2008}.

{While these developments have largely focused on reconnection in isolated or idealised current sheets, many astrophysical plasmas are instead in a turbulent state, where reconnection may occur ubiquitously across a broad range of scales. In this context, \cite{Galishnikova2022} identified the parameter regime in which tearing-mediated reconnection operates across a range of scales in \emph{forced} MHD turbulence. Reconnection is also believed to operate in \emph{unforced}, non-helical MHD turbulence \citep{Bhat2021}, where it may additionally regulate the decay timescale of the turbulence, in conjunction with the conservation of some dynamically important quantity.}

{Two quantities have been proposed as candidates for this role. The first is the so-called Hosking integral \citep{Hosking2021,Hosking2023,Zhou+22,Brandenburg2023,Brandenburg2023a,Brandenburg2024}, whose conservation and gauge invariance were recently examined under broad classes of gauge choices by \cite{Hew2026}. The second is the mean-squared vector potential, or anastrophy \citep{Zhou2019,Bhat2021,Dwivedi2024,Anandavijayan2026}. To address the gauge dependence of anastrophy, \cite{Brandenburg+15} defined it instead as $\bra{A_\mathrm{2D}^2}$, where $A_\mathrm{2D}$ is the component of the magnetic vector potential along the intermediate eigenvector of the rate-of-strain tensor. This choice minimises the anastrophy but does not render it fully gauge-independent; see table~5 of the supplemental material of \cite{Brandenburg+15}.}

If reconnection indeed governs the decay of magnetic energy in such systems, its rate should imprint directly on the global decay law. However, numerical evidence presents a more nuanced picture. In particular, \cite{Brandenburg2024} reported a decay timescale at large Lundquist numbers that appears largely independent of the magnetic Prandtl number, ${Pr_\mathrm{M}}$ -- {the ratio of the plasma viscosity, $\nu$, to the resistivity, $\eta$}. This finding is in tension with expectations from plasmoid-mediated reconnection theory. For instance, \cite{Comisso2015} obtain a reconnection rate scaling as $\sim {Pr_\mathrm{M}}^{-1/2}$ in the high-$S$ regime, at high $Pr_\mathrm{M}$.
This scaling was obtained from analyses of the boundary-driven Taylor problem, where a current sheet is externally forced to form. 

The discrepancy raises a key question: does reconnection, as realised in dynamically evolving turbulent systems, actually follow the same scaling laws as in externally driven configurations? The Taylor problem represents a highly idealised setup, in which the current sheet is both formed and driven by imposed boundary perturbations. In such studies \citep[e.g.,][hereafter CGW15]{Comisso2015}, the system typically begins from a current sheet that is initially stable to tearing, and reconnection is triggered by externally imposed perturbations. In contrast, current sheets in decaying turbulence arise self-consistently from the non-linear interaction and coalescence of magnetic structures, which may be spontaneously unstable to the plasmoid instability.

Motivated by this distinction, we revisit the $Pr_\mathrm{M}$ dependence of the reconnection rate using a minimal, self-consistent setup that resembles the magnetic structures observed in simulations of (at least two-dimensional) decaying MHD turbulence. Specifically, we study reconnection in the current sheet that sits at the interface of two coalescing magnetic islands. This configuration captures the spontaneous generation of plasmoids within a pre-existing current sheet during island coalescence, as opposed to externally driven setups.

Our study complements previous theoretical and numerical investigations of viscous magnetic reconnection.
These include the visco-resistive Sweet-Parker model of \citet{Park1984}, the plasmoid-mediated reconnection theory of \citet{Comisso2015}, simulations incorporating viscosity along with thermal conduction \citep{Minoshima2016}, and simulations of tearing-mode evolution with anisotropic (Braginskii) viscosity \citep{Tenerani2015}.

This paper is organised as follows. Section~\ref{sec:Numerical_Setup} presents the numerical setup, including the governing equations, the initial configuration representing the two coalescing magnetic islands, and the diagnostics used to measure the reconnection rate. In section~\ref{sec:Results_from_DNS}, we present results from our direct numerical simulations (DNSs), which support only a weak $Pr_\mathrm{M}$ dependence of the reconnection rate in the plasmoid-mediated regime.
In section~\ref{sec:taylor}, we compare these results with simulations of the boundary-driven Taylor problem. We also discuss the key physical differences between the two setups that may underlie the different $Pr_\mathrm{M}$ scalings obtained. Finally, in section~\ref{sec:discussion}, we summarise our findings and discuss their possible implications.

\section{Numerical Setup}
\label{sec:Numerical_Setup}

We perform 2D direct numerical simulations of a system of two
coalescing islands spanning a wide range of Lundquist and magnetic 
Prandtl numbers. 

\subsection{Basic equations}
We use the \textsc{Pencil Code} \citep{PC} to simulate the 
compressible isothermal visco-resistive equations of magnetohydrodynamics,

\begin{equation}
\frac{\partial\AAA}{\partial t}=\UU\times\BB+\eta\nabla^2\AAA,
\label{dAdt}
\end{equation}
\begin{eqnarray}
\frac{\DD\UU}{\DD t}=-\cs^2\nab\ln\rho
+\frac{1}{\rho}\left[\JJ\times\BB+\nab\cdot(2\rho\nu\SSSS)\right],
\label{dUdt}
\end{eqnarray}
\begin{equation}
\frac{\DD\ln\rho}{\DD t}=-\nab\cdot\UU,
\end{equation}
where 
$\DD/\DD t=\partial/\partial t+\UU\cdot\nab$ is the advective derivative, 
$\BB=\nab\times\AAA$ is the magnetic field 
(again in units where the vacuum permeability is unity),
$\JJ=\nab\times\BB$ is the current density,
$\UU$ is the velocity field,
$c_s$ is the sound speed,
$\nu$ is the viscosity, 
$\eta$ is the resistivity, and
${\sf S}_{ij}=(\partial_i U_j+\partial_j U_i)/2-\delta_{ij}\nab\cdot\UU/3$
are the components of the traceless rate-of-strain tensor $\SSSS$.
For a subset of runs, we use a GPU-accelerated version of the 
\textsc{Pencil Code} coupled with the \textsc{Astaroth} library
\citep{Pekkila2022},
enabling efficient simulations at higher resolutions.

The computational domain is a square box with sides
$L_x = L_y = 1$, 
spanning $-\left({L_x - \Delta x}\right)/{2} \le x \le \left({L_x + \Delta x}\right)/{2}$
and $-\left({L_y - \Delta y}\right)/{2} \le y \le \left({L_y + \Delta y}\right)/{2}$,
where $\Delta x = L_x/(N-1)$ and $\Delta y = L_y/(N-1)$ are the grid 
spacings in the $x$ and $y$ directions, respectively, and $N$ is 
the number of grid points in each direction. 
This choice ensures that there exist grid points that lie exactly 
at $x=0$ and $y=0$. 
Conducting, free-slip boundary conditions are imposed on all 
boundaries.

\subsection{Initial conditions}
\label{InitialConditions}

We adopt the initial configuration of two coalescing islands as 
described in \cite{Huang2010} (HB10 hereafter) for our numerical experiments. 
We initialise the system with no flows and a magnetic vector potential such 
that only the $z$-component is non-zero,
\begin{equation}
A_z(x,y) = {A_\mathrm{amp}}\tanh\!\left(\frac{x}{a}\right)
\cos\!\left(\frac{\pi y}{L_y}\right)
\sin\!\left(\frac{2\pi x}{L_x}\right),
\end{equation}
where $A_\mathrm{amp}$ sets the amplitude and $a$ is a measure of the 
current sheet half-width. 
The magnetic field is obtained from 
$\BB = \nabla \times \AAA$.
\begin{figure}
    \centering
    \includegraphics[width=0.75\linewidth]{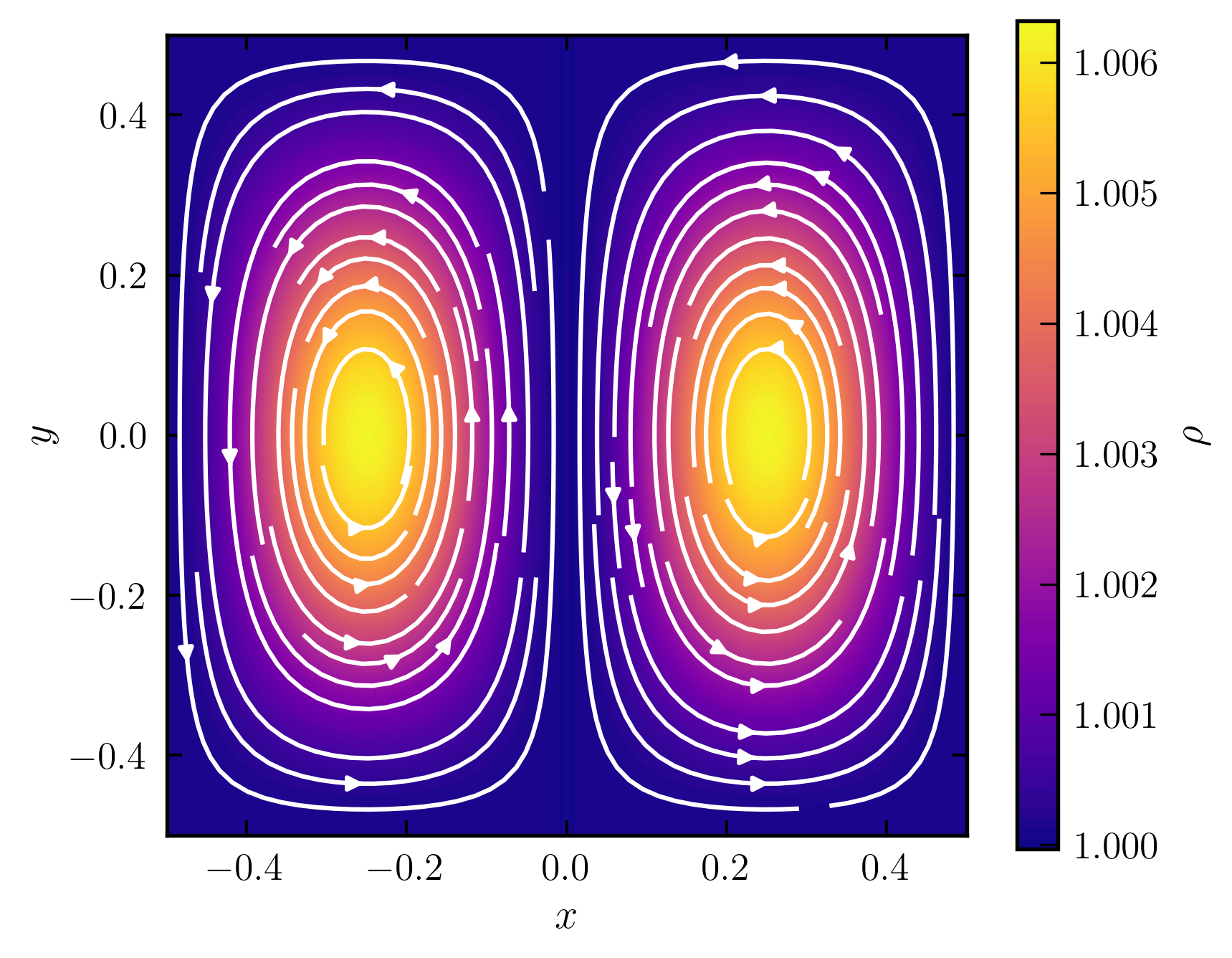}
    \caption{The initial configuration showing the two coalescing islands at {$t=0$}. The colour map represents the density, while the magnetic field is visualised using streamlines of $\BB$. This particular snapshot is from a run with $S=40000$ and the corresponding current sheet half-width, {$a=0.005$}.}
    \label{fig:rho_B_stream}
\end{figure}

We choose $A_\mathrm{amp} = 1/{2\pi}$ so that the resulting reconnecting field 
has an amplitude $B_\mathrm{amp} = 1$. The streamlines in \Fig{fig:rho_B_stream} show the magnetic field structure. The resulting current sheet is extended along the $y$-direction with the current density pointing along the negative $z$ direction.

This configuration is a smooth 
regularisation of
\begin{equation}
A_{z,0}(x,y) = {A_\mathrm{amp}}
\cos\!\left(\frac{\pi y}{L_y}\right)
\sin\!\left(\frac{2\pi |x|}{L_x}\right), \, 
A_{y,0} = A_{x,0} = 0\, .
\end{equation}
where the discontinuity at $x=0$ is replaced by a tanh profile,
yielding a current sheet of half-width $a$.
We choose the current sheet half-width to have a 
classical Sweet-Parker scaling such that $a = L_yS^{-1/2}$. 

The density is initialised to ensure an approximate force balanced 
equilibrium,
\begin{equation}
\rho(x,y) = \rho_0
+ \frac{5}{4} A_{z,0}^2\,
\frac{(2\pi/L_x)(2\pi/L_y)}{2c_s^2}
+ \frac{B_{y,0}^2 - B_y^2}{2c_s^2},
\end{equation}
where $\rho_0=1$, and we set $\ln \rho$ accordingly. 
This density profile is shown in \Fig{fig:rho_B_stream}.
Here, $B_y =  \BB \cdot \yyy$ and $B_{y,0} = (\nabla \times \AAA_0)\cdot \yyy$.
Note here that this is not simply a pressure equilibrium, but the
non-uniform density approximately balances the full Lorentz force,
the sum total of the magnetic tension and the magnetic pressure.
This is an important subtlety that we discuss in Appendix~\ref{app:PressEq}. 

The important non-dimensional number that we characterise the
system with is the Lundquist number $S=V_AL/\eta$. 
The characteristic length scale for us is the length of the 
current sheet, $L_y=1$, the Alfv\'en speed, 
$V_A = B_\mathrm{amp}/\sqrt{\rho_0} = 1$, 
and so the Lundquist number is simply $S=1/\eta$.
Further, we minimise compressibility effects by setting the sound speed to $10 V_A$ which implies a plasma beta of $\beta = \rho_0c_s^2/2B^2= 50$. The Alfv\'en time is defined as the ratio of the length of the current sheet to the Alfv\'en speed, $\tau_A = L/V_A = 1$.

{The other relevant dimensionless parameters are the magnetic Reynolds number, $Re_\mathrm{M} = UL/\eta$, the kinetic Reynolds number, $Re = UL/\nu$, and the magnetic Prandtl number, $Pr_\mathrm{M} = \nu/\eta$. Since reconnection outflows are expected to be Alfv\'enic, we take the characteristic velocity to be the Alfv\'en speed, $U = V_A$. With the current-sheet length $L=L_y=1$, this choice gives $Re_\mathrm{M} = V_AL/\eta = S$, so the magnetic Reynolds and Lundquist numbers are identical in our setup.}

\subsection{Diagnostics}
\label{Diagnostics}

The primary quantity we aim to estimate is the reconnection rate. 
We adopt a measure similar to that proposed by HB10.
The instantaneous non-dimensional reconnection rate is defined through the rate of depletion of magnetic flux,
\begin{equation}
V_{\mathrm{rec}} = -\dfrac{1}{V_A\,B_\mathrm{amp}}\dfrac{d}{dt}
\left[
\max_{x} A_z(x,0,t) \;-\; \max_{y} A_z(0,y,t)
\right].
\label{eq:rec-rate}
\end{equation}
Here, the quantity inside the brackets measures the magnetic flux contained within an island,
$\Phi_{\mathrm{island}}$ (see Appendix~\ref{app:Rates} for further details).
As reconnection proceeds, this flux decreases, and its time derivative therefore provides a direct measure of the reconnection rate.
The definition in \Eq{eq:rec-rate} is identical to that used by HB10 
(see also \citealt{Comisso2015} for a closely related diagnostic), 
except that HB10 assume
\begin{equation}
\dfrac{d}{dt} \left[\max_{x} A_z(x,0,t)\right] = 0,
\end{equation}
{which is strictly valid only in the ideal ($\eta=0$) limit. At finite resistivity, $\max_x A_z(x,0,t)$ evolves solely owing to resistive diffusion, whereas $\max_y A_z(0,y,t)$ evolves owing to both resistive diffusion and magnetic reconnection. Neglecting the evolution of $\max_x A_z(x,0,t)$ therefore attributes not only the reconnection-driven flux depletion but also the diffusive component of the flux depletion to reconnection, leading to a slight overestimate of the reconnection rate. Retaining this term removes the contribution from resistive diffusion, thereby isolating the reconnection-driven flux depletion and providing a more accurate estimate of the reconnection rate.}
An alternative measure of the reconnection rate used by \cite{Vicentin2026} is described in Appendix~\ref{app:Rates}, and gives consistent results.

\section{Results from direct numerical simulations}
\label{sec:Results_from_DNS}

\begin{table}
\centering
\small
\setlength{\tabcolsep}{5pt}

\begin{tabular}{cccccc}
\hline
S.No & Resolution, {$N$} & $S \times 10^{3}$ & ${Pr_\mathrm{M}}$ &  {Current sheet half-width,} $a$ \\
\hline
1  & 1024  & 1   & 1, 10, 20, 40      & 0.0316 \\
2  & 1024  & 2   & 1                  & 0.0224 \\
3  & 1024  & 4   & 1, 10, 20, 40      & 0.0158 \\
4  & 4096  & 8   & 1                  & 0.0112 \\
5  & 4096  & 10  & 1                  & 0.0100 \\
6  & 4096  & 20  & 1                  & 0.0071 \\
7  & 4096  & 40  & 1, 10, 20, 40, 100 & 0.0050 \\
8  & 4096  & 100 & 1                  & 0.0032 \\
9* & 6144  & 100 & 1                  & 0.0032 \\
10 & 16384 & 200 & 1                  & 0.0022 \\
11 & 16384 & 500 & 1, 10, 20, 40, 50  & 0.0014 \\
12* & 32768 & 500 & 1                 & 0.0014 \\
\hline
\end{tabular}
\caption{{Summary of the simulation parameters. The columns give the numerical resolution $N$, Lundquist number $S$, magnetic Prandtl number $Pr_\mathrm{M}$, and the initial current-sheet half-width $a$. Runs marked with an asterisk (*) denote the high-resolution simulations used for the convergence study (see Appendix~\ref{app:convergence}).}}
\label{tab:sim_params}
\end{table}
In this section, we present results from a suite of direct
numerical simulations performed over a range of Lundquist numbers 
and magnetic Prandtl numbers. 

\subsection{Runs with $Pr_\mathrm{M}=1$}

We first summarise results from $Pr_\mathrm{M}=1$ simulations to check 
consistency with existing literature 
and recover known scaling relations. \TTab{tab:sim_params} summarises various run parameters for all the runs.

\begin{figure}
    \centering
    \includegraphics[width=0.75\linewidth]{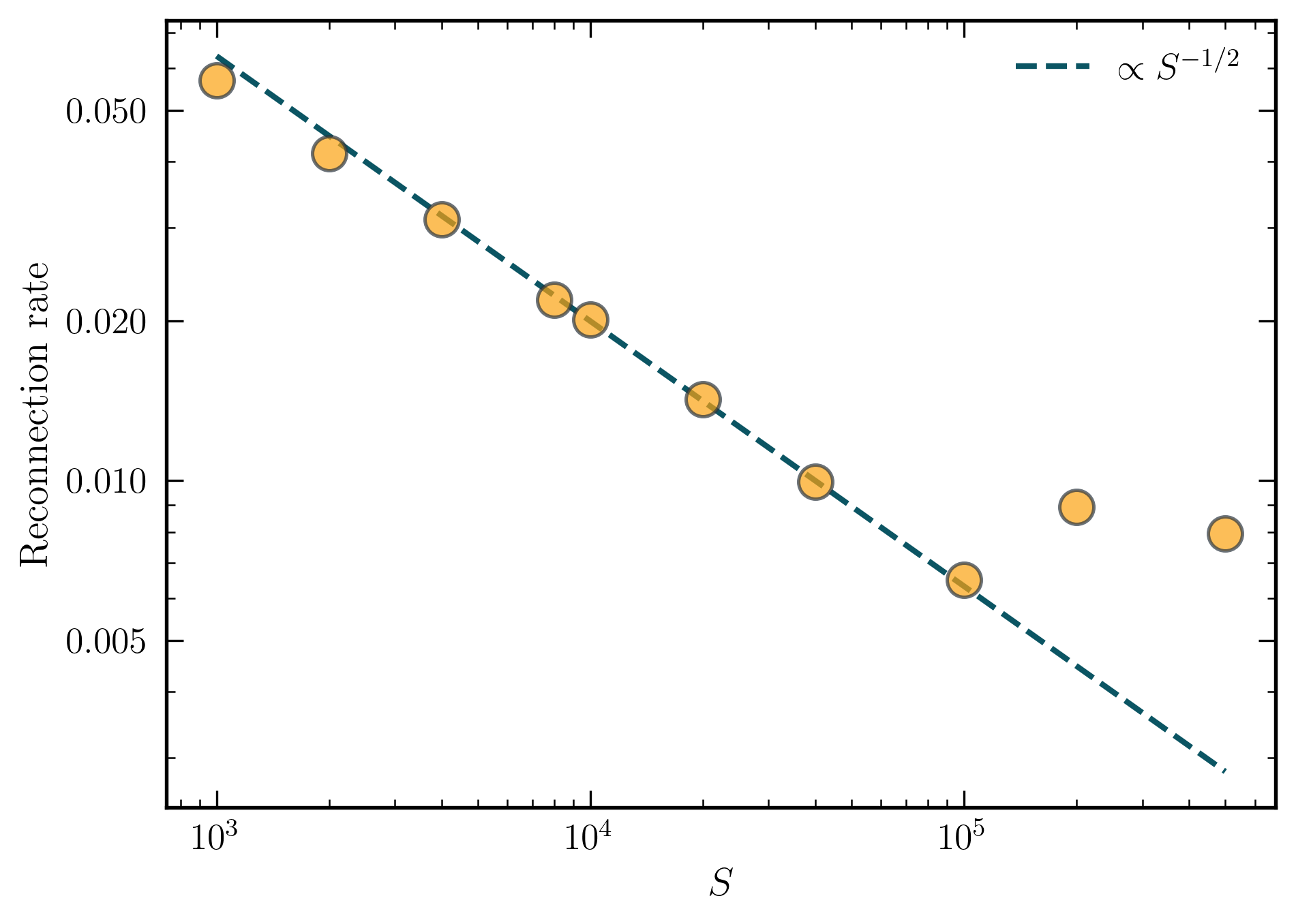}
    \caption{The scaling of the reconnection rate vs Lundquist number, $S$ at $Pr_{\rm M} = 1$. The orange circles show the reconnection rates obtained from simulations. The black dashed line shows the Sweet-Parker scaling, $V_\mathrm{rec} \propto S^{-1/2}$.}
    \label{fig:reconnection_rate_scaling_S}
\end{figure}

\FFig{fig:reconnection_rate_scaling_S} shows the dependence of this non-dimensionalised reconnection rate on the Lundquist number, $S$.
Consistent with existing results, we find that the non-dimensionalised reconnection rate follows the Sweet-Parker scaling $V_\mathrm{rec} \propto S^{-1/2}$ up to a critical Lundquist number, $S_c \approx 10^5$. Close to and beyond $S_c$ the plasmoid instability sets in, giving rise to a Lundquist number independent regime. {Unlike recent work by \cite{Vicentin2026} with a similar setup, we do not observe an intermediate $V_\mathrm{rec}\propto S^{-1/3}$ scaling. In Appendix~\ref{app:PressEq}, we show that this scaling can nevertheless be recovered using their initial condition, suggesting that it is sensitive to the initial force balance rather than representing a universal tearing-mediated reconnection regime.}

\subsection{$Pr_\mathrm{M}$ dependence for $S<S_c$}

{
In the Sweet-Parker regime, the dependence of the reconnection rate on the magnetic Prandtl number arises from the influence of viscosity on the outflow from the current sheet. \citet{Park1984} first showed that viscous stresses reduce the outflow velocity, thereby broadening the current sheet through mass conservation and reducing the inflow speed; see also the modern discussion by \citet{Baty2022}. We briefly summarize this scaling argument before comparing it with the reconnection rates measured in our simulations.
}

 \subsubsection{Visco-resistive Sweet-Parker scaling}
 {
For a steady Sweet-Parker current sheet of length $L_{\rm SP}$ and thickness
$\delta_{\rm SP}$, with inflow speed $V_{\rm in}$ and outflow speed
$V_{\rm out}$, mass conservation gives
\begin{equation}
V_{\rm in}L_{\rm SP}
\sim
V_{\rm out}\delta_{\rm SP}.
\label{eq:mass_balance}
\end{equation}
Balancing advection and diffusion in the steady induction equation,
\begin{equation}
\frac{V_{\rm in}B}{\delta_{\rm SP}}
\sim
\eta\frac{B}{\delta_{\rm SP}^2},
\end{equation}
yields
\begin{equation}
V_{\rm in}
\sim
\frac{\eta}{\delta_{\rm SP}}.
\end{equation}
Combining these relations gives
\begin{equation}
\delta_{\rm SP}^2
\sim
\frac{\eta L_{\rm SP}}{V_{\rm out}}.
\label{eq:delta_balance}
\end{equation}

The outflow speed follows from an energy balance. The magnetic energy carried
into the current sheet per unit time,
\begin{equation}
P_{\rm mag}
\sim
\frac{B^2}{2}\,
V_{\rm in}L_{\rm SP},
\end{equation}
is converted into kinetic energy of the exhaust together with viscous
dissipation inside the current sheet.

The kinetic energy flux leaving the two exhausts scales as
\begin{equation}
P_{\rm kin}
\sim
\rho V_{\rm out}^3\delta_{\rm SP},
\end{equation}
while the viscous dissipation rate is
\begin{equation}
P_\nu
\sim
C_\nu \rho\nu
\left(\frac{V_{\rm out}}{\delta_{\rm SP}}\right)^2
L_{\rm SP}\delta_{\rm SP}
=
C_\nu \rho\nu
\frac{V_{\rm out}^2L_{\rm SP}}
{\delta_{\rm SP}},
\end{equation}
where $C_\nu$ is a dimensionless coefficient that
depends on the detailed velocity profile across the current sheet.
Using Eq.~(\ref{eq:mass_balance}) to eliminate $V_{\rm in}$, the energy balance
\begin{equation}
P_{\rm mag}
\sim
P_{\rm kin}
+
P_\nu,
\end{equation}
becomes
\begin{equation}
V_{\rm out}^2
+
C_\nu
\frac{\nu L_{\rm SP}}
{\delta_{\rm SP}^2}
V_{\rm out}
\sim
V_A^2,
\label{eq:energy_balance}
\end{equation}
where $V_A=B/\sqrt{\rho}$.
Using Eq.~(\ref{eq:delta_balance}) gives
\begin{equation}
V_{\rm out}^2
\left(
1+C_\nu\frac{\nu}{\eta}
\right)
\sim
V_A^2,
\end{equation}
or
\begin{equation}
V_{\rm out}
\sim
\frac{V_A}
{\sqrt{1+C_\nu Pr_{\rm M}}}.
\label{eq:vout_general}
\end{equation}

Substituting this into Eq.~(\ref{eq:delta_balance}) yields
\begin{equation}
\delta_{\rm SP}
\sim
L_{\rm SP}
S^{-1/2}
(1+C_\nu Pr_{\rm M})^{1/4},
\end{equation}
and therefore
\begin{equation}
V_{\rm rec}
=
\frac{V_{\rm in}}{V_A}
=
\frac{V_{\rm out}}{V_A}
\frac{\delta_{\rm SP}}{L_{\rm SP}}
\sim
S^{-1/2}
(1+C_\nu Pr_{\rm M})^{-1/4}.
\label{eq:park_general}
\end{equation}
Factoring out the coefficient $C_\nu$,
\begin{equation}
1+C_\nu Pr_{\rm M}
=
C_\nu
\left(
Pr_{\rm M}
+\frac{1}{C_\nu}
\right),
\end{equation}
defining
$Pr_{{\rm M}0}\equiv C_\nu^{-1}$,
and absorbing the constant factor $C_\nu^{-1/4}$ into the overall
proportionality coefficient gives
\begin{equation}
V_{\rm rec}
\sim
S^{-1/2}
(Pr_{{\rm M}0}+Pr_{\rm M})^{-1/4},
\label{eq:park_shifted}
\end{equation}
which reduces to
\begin{equation}
V_{\rm rec}
\propto
S^{-1/2}
Pr_{\rm M}^{-1/4},
\qquad
Pr_{\rm M}\gg1.
\end{equation}
}

\subsubsection{Comparison with simulations}

{We now compare the above scaling with the reconnection rates measured in our simulations in the Sweet-Parker ($S<S_c$) regime.}

\begin{figure}
    \centering
    \includegraphics[width=\linewidth]{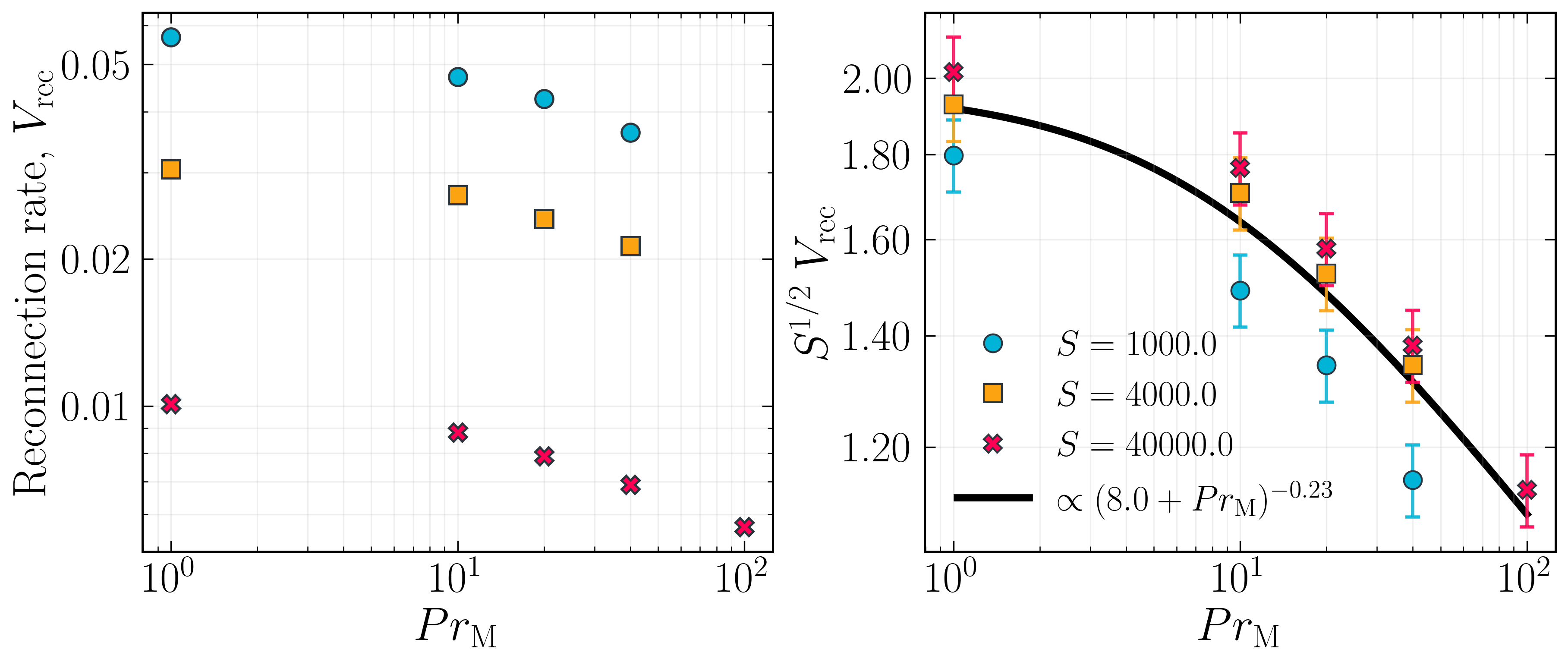}
    \caption{
    {Dependence of the reconnection rate on the magnetic Prandtl number, $Pr_{\rm M}$, for three Lundquist numbers, $S=1000,\ 4000,$ and $40000$ in the Sweet-Parker regime. \textit{Left:} Measured quasi-steady reconnection rate, $V_{\rm rec}$, as a function of $Pr_{\rm M}$. For all three Lundquist numbers, the reconnection rate decreases with increasing magnetic Prandtl number. \textit{Right:} Reconnection rates rescaled by $S^{1/2}$, showing an approximate collapse of the data onto a single curve. The error bars denote an assumed uncertainty of $5\%$. The solid curve shows the best-fitting model, $S^{1/2}V_{\rm rec} =  1.97(1.0+Pr_{\rm M}/8.0)^{-0.23}$, indicating that the rescaled reconnection rate is well described by the theoretical prediction in \Eq{eq:park_shifted}}.
    }   
    \label{fig:V_rec_vs_Pm_lowS}
\end{figure}

{\FFig{fig:V_rec_vs_Pm_lowS} compares the measured reconnection rates with the visco-resistive Sweet-Parker scaling for three Lundquist numbers, $S=1000,\ 4000,$ and $40000$. The left panel shows that, for each value of $S$, the quasi-steady reconnection rate decreases with increasing magnetic Prandtl number. At the same time, the reconnection rate decreases with {increasing} Lundquist number, consistent with the Sweet-Parker scaling $V_{\rm rec}\propto S^{-1/2}$.}

\begin{figure}
    \centering
    \includegraphics[width=1\linewidth]{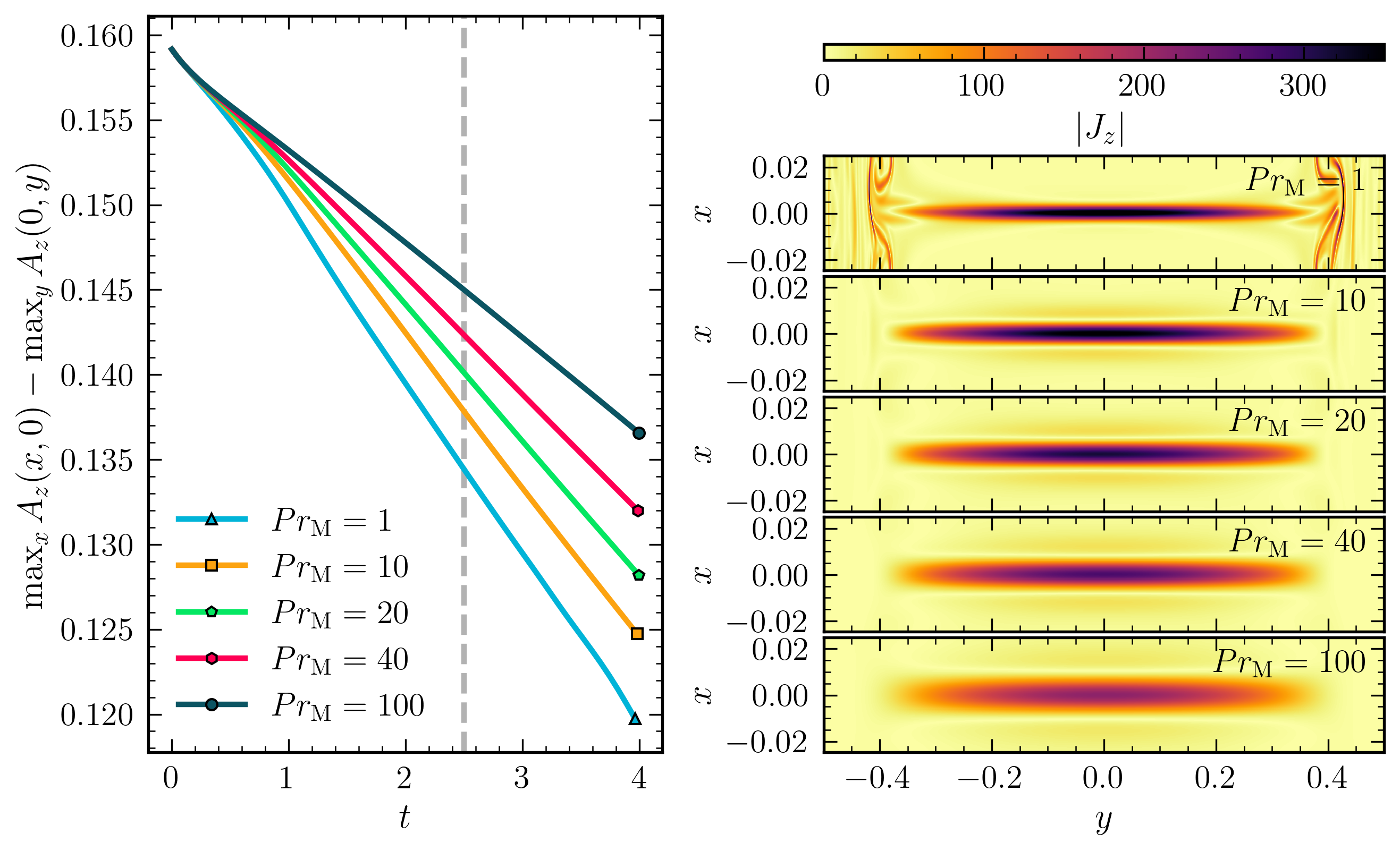}
    \caption{Representative current-density maps ($|J_z|$) for the Sweet-Parker regime at $S=40000$ and magnetic Prandtl numbers $Pr_\mathrm{M}=1$, 10, 20, 40, and 100. The left panel shows the evolution of the magnetic flux contained within the islands, with the vertical dashed line indicating the time at which the current-density snapshots are shown in the right panels.}
    \label{fig:SPsheets}
\end{figure}

{To isolate the dependence on $Pr_{\rm M}$, the right panel shows the reconnection rates rescaled by $S^{1/2}$. The rescaled data exhibit an approximate collapse onto a single curve, particularly for the higher Lundquist numbers, consistent with the expected $S^{-1/2}$ scaling over the range of $S$ considered. We fit the rescaled reconnection rates with the shifted power-law
\begin{equation}
V_{\rm rec}S^{1/2} = V_0 (1+Pr_{\rm M}/Pr_{{\rm M}0})^\alpha,
\end{equation}
obtaining $V_0 = 1.97$, $Pr_{{\rm M}0}=8.0$ and $\alpha=-0.23$, in good agreement with the theoretical prediction ($\alpha = -0.25$) of \Eq{eq:park_shifted}. No plasmoid formation is observed in any of these simulations, confirming that the current sheet remains in the laminar Sweet-Parker regime throughout the measurement interval. This is illustrated in \Fig{fig:SPsheets}, which shows representative current-density maps for the $S=40000$ runs spanning the range of $Pr_\mathrm{M}$ considered. In all cases the current sheet remains smooth and elongated, with no evidence of secondary islands or plasmoid formation.}

\subsection{$Pr_\mathrm{M}$ dependence for the plasmoid regime}
\label{subsec:Pm-dep-plasmoid}

\begin{figure}
    \centering
    \includegraphics[width=1.\linewidth]{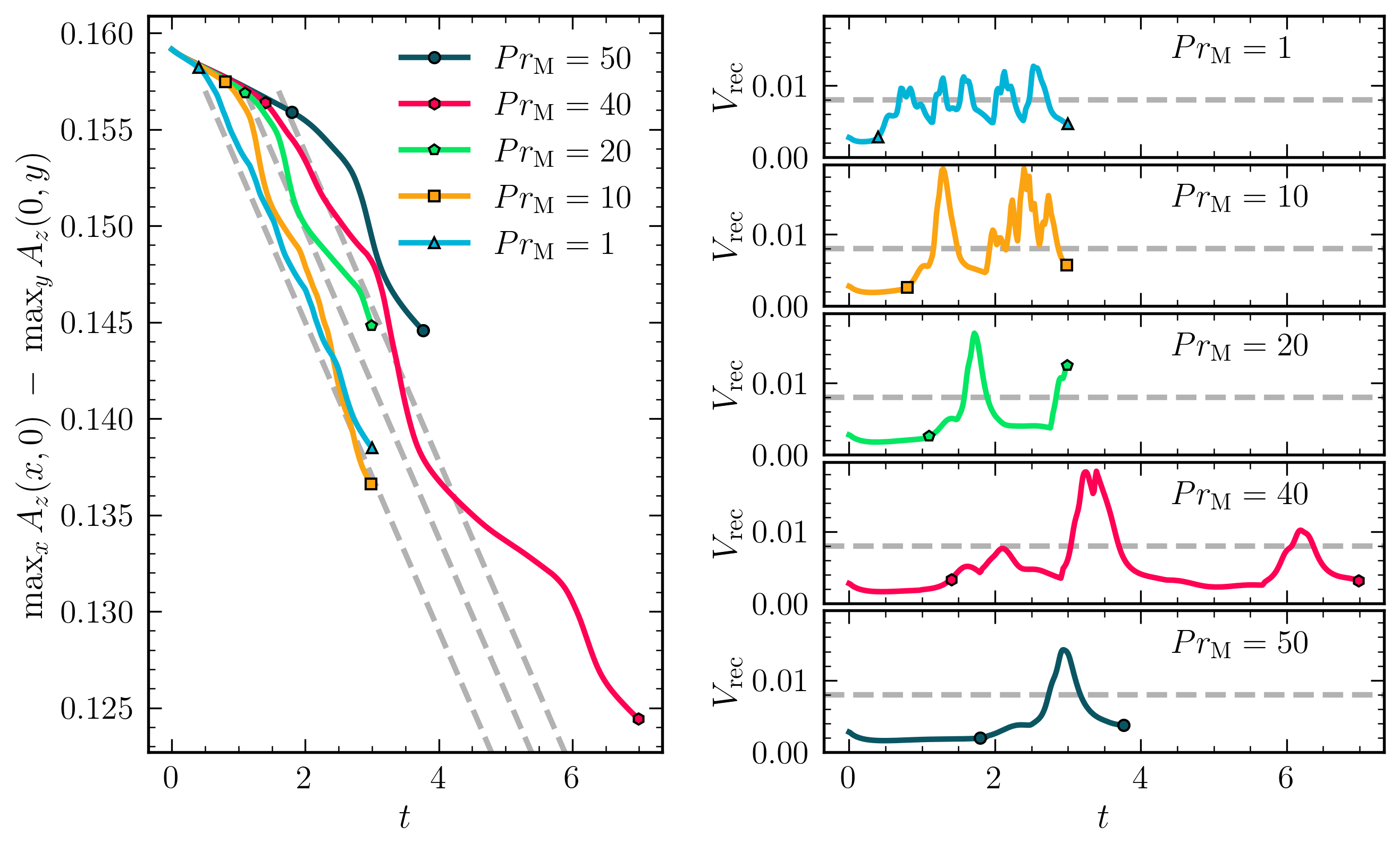}
    \caption{Time evolution of the magnetic flux and the corresponding reconnection rate for different magnetic Prandtl numbers at $S=5\times10^5$.
    The left panel shows the quantity
    $\max_x A_z(x,0)-\max_y A_z(0,y)$,
    which measures the magnetic flux contained within the magnetic island.
    The right panel shows the corresponding reconnection rate, computed from the time derivative of this quantity as defined in equation~(\ref{eq:rec-rate}). The dashed lines indicate a reconnection rate of $0.008$, obtained from a linear fit to the flux evolution in the $Pr_\mathrm{M}=1$ run. {The paired markers indicate the time interval used to compute the mean reconnection rate and its $1\sigma$ uncertainty for each simulation. The start of the averaging interval was selected by visual inspection as the point where the reconnection rate increases following the initial transient, while the end coincides with the end of the simulation.}}
    \label{fig:S5e5_PmDep}
\end{figure}

We now turn to the regime in which the current sheet becomes unstable to the plasmoid instability and fragments into multiple secondary islands. In our simulations, this transition is clearly observed for $S \gtrsim 10^5$. {Previous studies have shown that viscosity influences not only the Sweet–Parker reconnection rate, but also the linear tearing instability itself by reducing its growth rate and increasing the characteristic tearing time, with the precise scaling depending on the tearing regime (e.g. \citet{Ofman1991}). In the present work, however, we are primarily interested in the reconnection rate after the current sheet has fully entered the plasmoid-mediated regime. The reconnection rates reported below are therefore measured well after the initial onset of the tearing instability, when the current sheet has fragmented into multiple plasmoids and the dynamics are dominated by their non-linear evolution and interactions.}

We focus on runs with $S = 5\times10^5$ and magnetic Prandtl numbers in the range $1 \leq Pr_\mathrm{M} \leq 50$. 
{All simulations were performed at a numerical resolution of $16384^2$. To assess numerical convergence, the $Pr_{\rm M}=1$ case was additionally repeated at twice the resolution, $32768^2$.}
The evolution of the magnetic flux contained within the islands, together with the corresponding reconnection rates, is shown in figure~\ref{fig:S5e5_PmDep}. 

We find that the reconnection rate exhibits only a weak dependence on $Pr_\mathrm{M}$ across this range, remaining broadly consistent with a $Pr_\mathrm{M}$-independent scaling. In particular, the instantaneous reconnection rate in all runs reaches, and often exceeds, values of $\approx 0.008$, consistent with the characteristic rates expected for plasmoid-mediated reconnection \citep{Bhattacharjee2009,Huang2010,Loureiro2012}.

In particular, let us focus on the run with $Pr_\mathrm{M}=40$.
As can be seen from figure~\ref{fig:S5e5_PmDep}, the reconnection proceeds in a quasi-cyclic manner,
alternating between phases of relatively fast and slow reconnection.
We investigate this behaviour in greater detail below and identify a plausible physical explanation for these oscillations.

\begin{figure}
    \centering
    \includegraphics[width=1\linewidth]{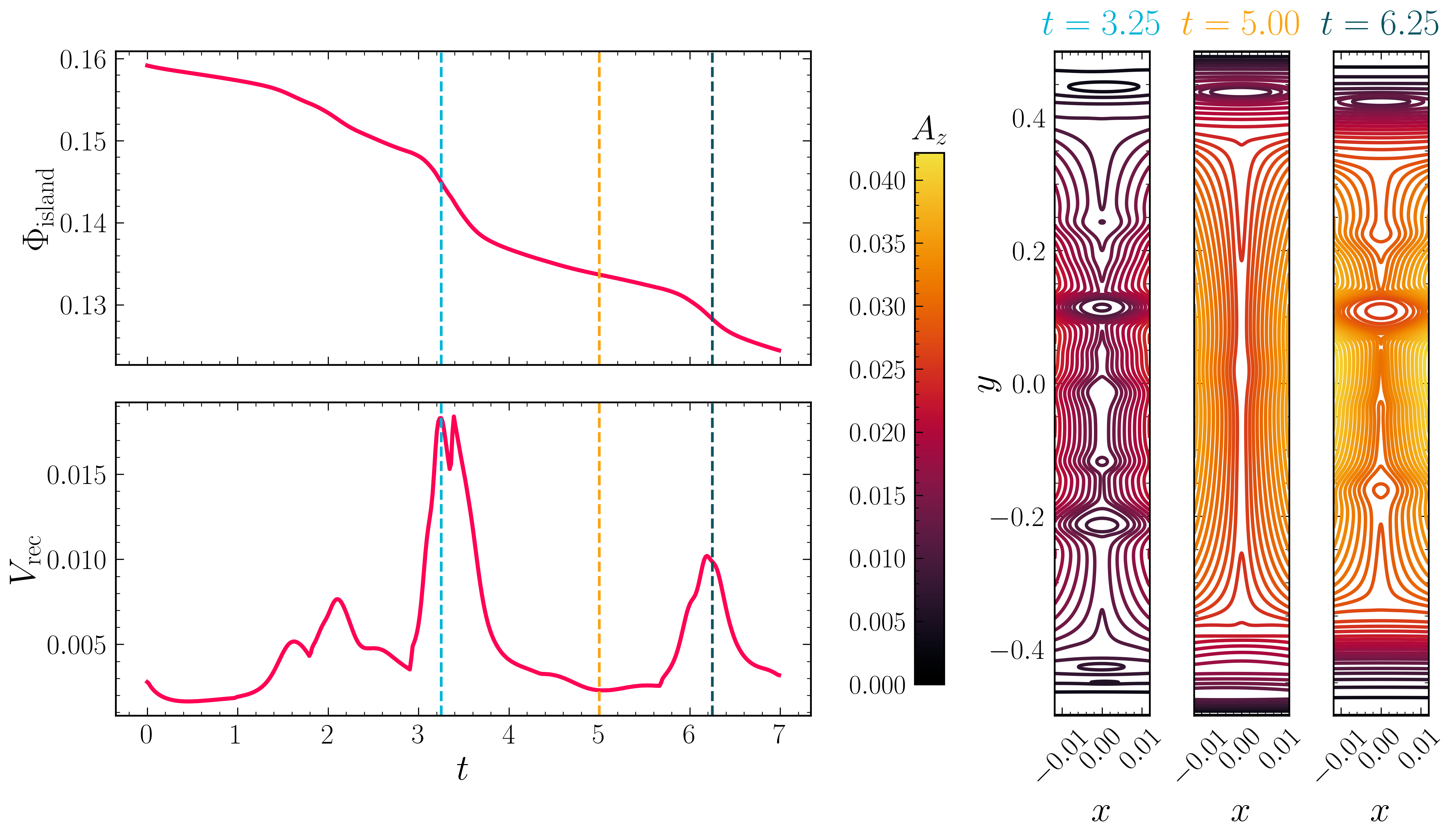}
    \caption{Time evolution of the magnetic flux $\Phi_{\mathrm{island}}$ (top left panel), and the corresponding reconnection rate, $V_{\mathrm{rec}}$ (bottom left panel), for the $S=5\times10^5$, $Pr_\mathrm{M}=40$ run in the plasmoid-unstable regime. The vertical dashed lines mark the three times at which contours of the vector potential $A_z$ are shown in the right-hand panels. These snapshots correspond to distinct phases of the reconnection dynamics: a period of rapid reconnection associated with plasmoid interactions and mergers ($t=3.25$), a relatively quiescent phase with slower reconnection ($t=5.00$), and a later episode of enhanced reconnection rate accompanied by plasmoid activity ($t=6.25$). The cyclic alternation between faster and slower reconnection phases is reflected in the temporal variability of $V_{\mathrm{rec}}$.}
    \label{fig:16384h_Pm40_rate_oscillations}
\end{figure}

We reproduce in figure~\ref{fig:16384h_Pm40_rate_oscillations} the flux evolution curve from figure~\ref{fig:S5e5_PmDep} for the run with $Pr_\mathrm{M} = 40$. Alongside this, we show contours of the vector potential at three representative times. Two of these snapshots ($t=3.25$ and $t=6.25$) correspond to phases of relatively fast reconnection ($V_\mathrm{rec}\gtrsim0.008$), while the intermediate snapshot ($t=5.00$) corresponds to a slower reconnection phase. The contour plots reveal a marked qualitative difference between these states. During the fast-reconnection phases, the current sheet contains multiple plasmoids of appreciable size, many of which appear to be interacting and undergoing mergers. In contrast, during the quiescent phase the layer is dominated by a single comparatively small plasmoid and exhibits much less dynamical activity. This suggests that interactions and coalescence between plasmoids may play an important role in sustaining enhanced reconnection rates. {A similar conclusion was reached by \citet{Vicentin2026}, who found that sustained fast reconnection is associated with a strongly non-linear regime characterized by vigorous inter-plasmoid interactions and mergers, rather than merely by the onset of plasmoid formation.}

Moreover, in the plasmoid-mediated regime, the number of plasmoids is expected to increase with the Lundquist number (see e.g., HB10). {We therefore hypothesise that, at astrophysically large Lundquist numbers, the current sheet may host a sufficiently large population of interacting plasmoids that prolonged quiescent intervals become increasingly unlikely. If so, reconnection could remain persistently in the non-linear, interaction-dominated state, leading to a fast, $Pr_\mathrm{M}$-independent reconnection rate. Testing this conjecture, however, requires simulations at substantially higher Lundquist numbers than those considered here.}

\begin{figure}
    \centering
    \includegraphics[width=0.8\linewidth]{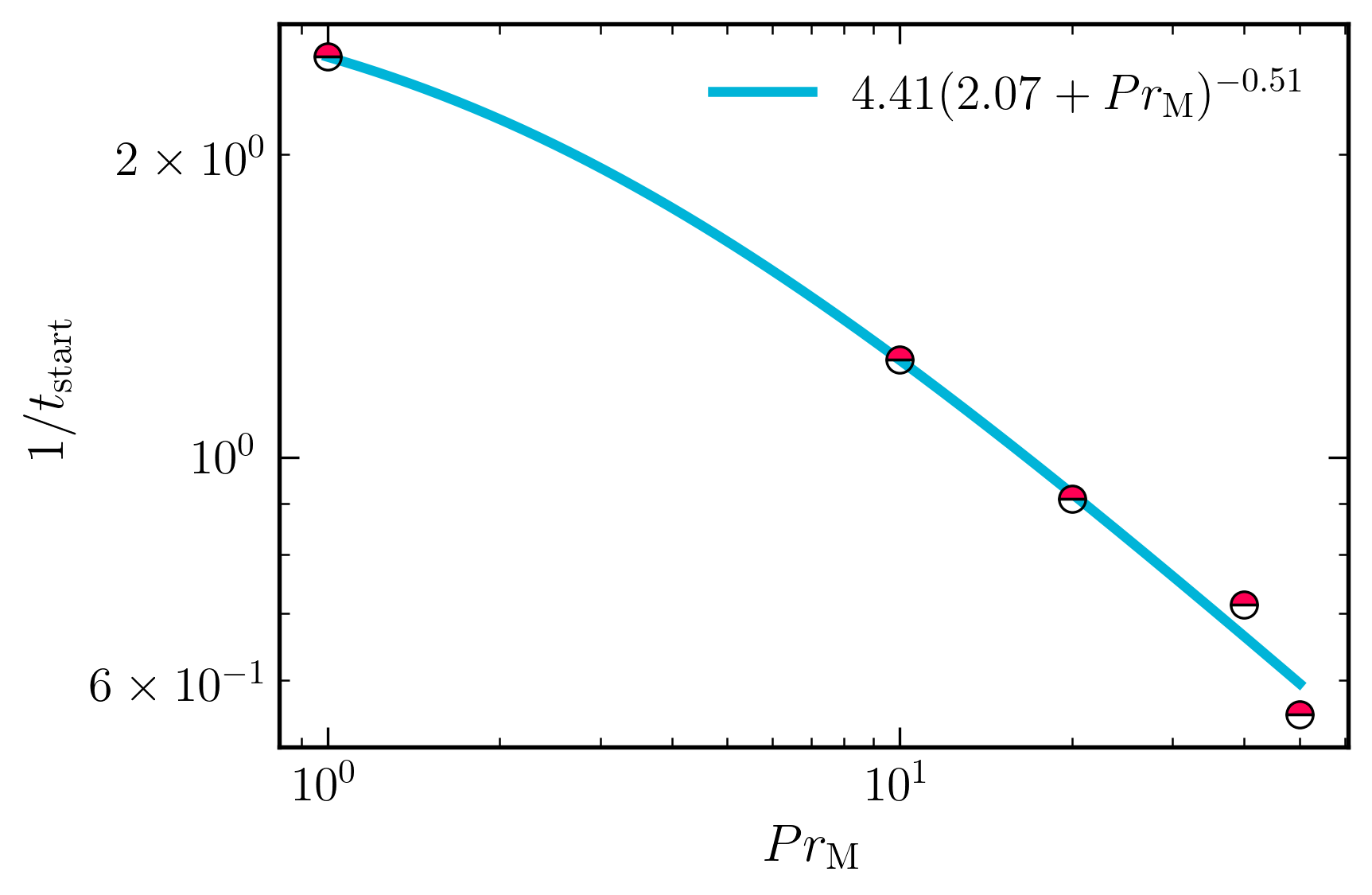}
    \caption{Inverse start time, $1/t_{\rm start}$, as a function of the magnetic Prandtl number, $Pr_{\rm M}$. The solid curve shows the best-fitting shifted power law, $1/t_{\rm start}=4.41\,(2.07+Pr_{\rm M})^{-0.51}$, consistent with the visco-resistive scaling of \citet{Comisso2015}.}
    \label{fig:tstart}
\end{figure}

To better quantify the dependence of the reconnection rate on the magnetic Prandtl number, we compute the time-averaged reconnection rate for each simulation over the intervals indicated by the paired markers in figure~\ref{fig:S5e5_PmDep}. The start time ($t_{\rm start}$) of each interval is chosen by visual inspection as the point where the reconnection rate begins to increase following the initial transient, while the end corresponds to the end of the simulation. 
Empirically, we find that the inverse onset time exhibits the same magnetic-Prandtl-number dependence as the visco-resistive plasmoid growth rate derived by \citet{Comisso2015}.
\begin{equation}
\frac{\tau_A}{t_{\rm start}}
=
4.41\left(2.07+Pr_{\rm M}\right)^{-0.51},
\label{eq:tstart_fit}
\end{equation}
as shown in \Fig{fig:tstart}. This can equivalently be written as
\begin{equation}
t_{\rm start}
=
t_0
\left(
1+\frac{Pr_{\rm M}}{2.07}
\right)^{0.51},
\label{eq:tstart_fit2}
\end{equation}
where
\begin{equation}
t_0
=
\frac{\tau_A}{4.41(2.07)^{-0.51}}
=
\frac{\tau_A}{3.42}
\approx
0.29\,\tau_A.
\end{equation}

At first sight, the fact that $t_0<\tau_A$ may appear surprising, since reconnection is often expected to occur on timescales longer than the Alfv\'en time. However, the relevant Alfv\'en timescale governing the tearing instability is not the global crossing time but the Alfv\'en time across the initial current-sheet half-thickness ($a$), given by
\begin{equation}
\tau_{A,a}
=
\frac{a}{V_A}
=
S^{-1/2}\tau_A.
\end{equation}
For the present simulations, $S=5\times10^5$, giving
\begin{equation}
\tau_{A,a}
=
S^{-1/2}\tau_A
\approx
1.41\times10^{-3}\tau_A.
\end{equation}
Consequently,
\begin{equation}
t_0
\approx
\frac{0.29}{1.41\times10^{-3}}
\,\tau_{A,a}
\approx
205\,\tau_{A,a},
\end{equation}
so that the ordering
\begin{equation}
\tau_{A,a}
\ll
t_0
<
\tau_A
\end{equation}
is entirely consistent with the expected hierarchy of timescales.

\begin{figure}
    \centering
    \includegraphics[width=0.9\linewidth]{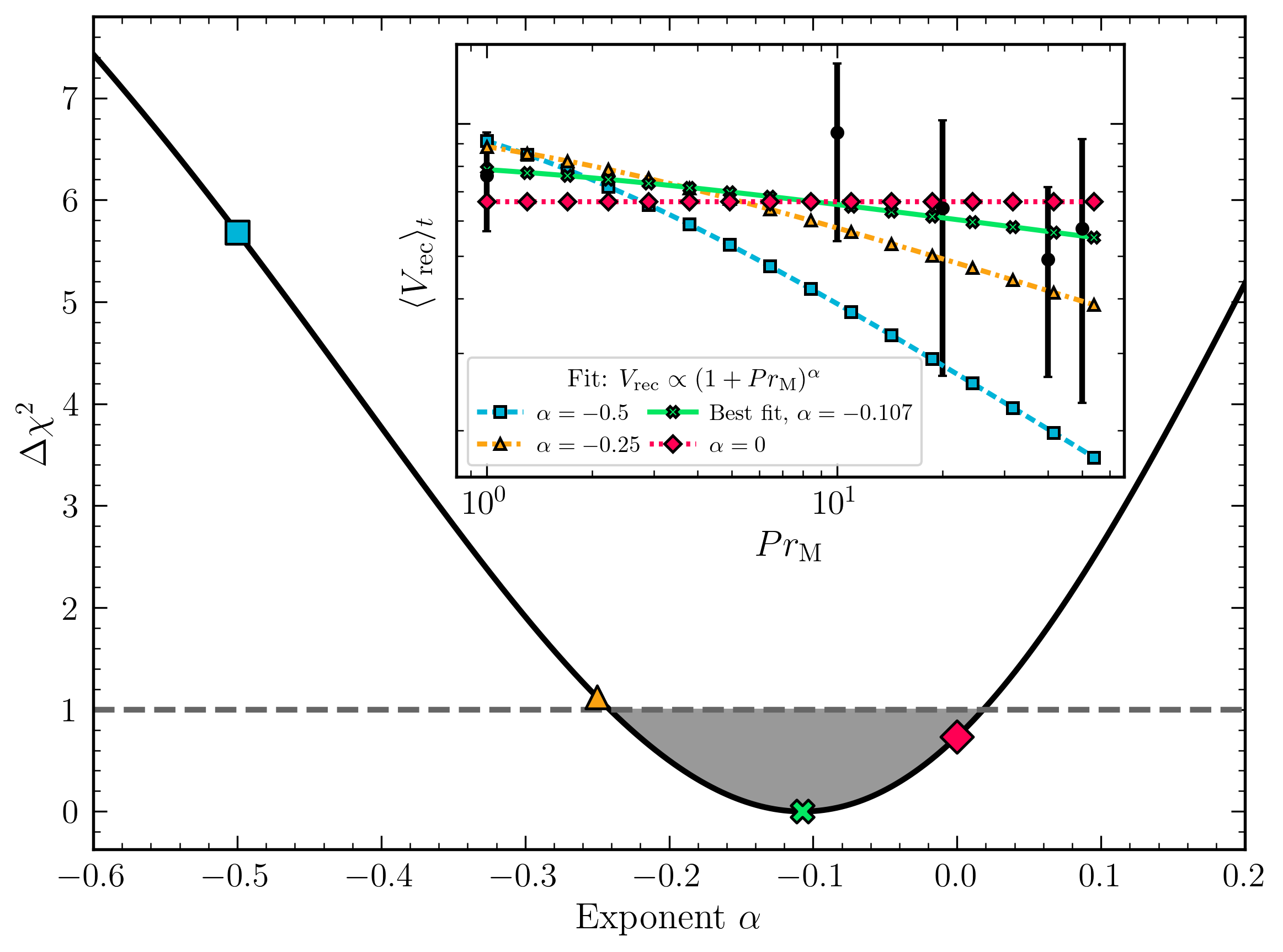}
    \caption{Constraints on the magnetic Prandtl number scaling of the time-averaged reconnection rate. The main panel shows $\Delta\chi^2$ for fits of $V_{\rm rec}=V_0(1+Pr_{\rm M})^\alpha$, with the dashed line and shaded region indicating the $68\%$ confidence interval. The inset compares the measured rates (circular markers with $1\sigma_{\rm mean}$ error bars) with fits for $\alpha=-1/2$, $-1/4$, $0$, and the best-fit exponent, $\alpha=-0.107$.}
    \label{fig:chi2_alpha}
\end{figure}

To estimate the uncertainty in the time-averaged reconnection rate, we approximately account for temporal correlations in the instantaneous reconnection rate by assuming a correlation time of one Alfvén time, $\tau_{\rm A}$, corresponding to the characteristic dynamical timescale of the system. Using the standard expression for correlated time series \citep{Sokal1997,Wolff2004}, we estimate the effective number of statistically independent samples as
\begin{equation}
N_{\rm eff}=\frac{T}{2\tau_{\rm A}},    
\end{equation}
where $T$ is the duration of the averaging interval. The uncertainty in the mean reconnection rate is then estimated as
\begin{equation}
\sigma_{\rm mean}=\frac{\sigma}{\sqrt{N_{\rm eff}}}, 
\end{equation}
where $\sigma$ is the standard deviation of the instantaneous reconnection rate over the averaging interval. 
The measured mean reconnection rates are then fitted using the functional form
\begin{equation}
V_{\rm rec}=V_0\left(1+Pr_{\rm M}\right)^\alpha,
\end{equation}
where both the normalisation $V_0$ and the exponent $\alpha$ are treated as free parameters \footnote{As in the previous sections, we also considered the three-parameter model
$
V_{\rm rec}=V_0\left(Pr_{\rm M0}+Pr_{\rm M}\right)^\alpha
$,
in which $V_0$, $\alpha$, and $Pr_{\rm M0}$ are all treated as free parameters.
A comparison of the resulting fits is presented in Appendix~\ref{app:fit_models}.}.
For each fixed value of $\alpha$, the corresponding best-fitting normalisation $V_0$ is determined by a weighted least-squares fit using the measured uncertainties, and the resulting $\chi^2$ is computed. Figure~\ref{fig:chi2_alpha} shows the corresponding $\Delta\chi^2=\chi^2-\chi^2_{\rm min}$ as a function of $\alpha$, together with representative fits to the measured reconnection rates in the inset. The minimum occurs at $\alpha=-0.107$, with a $68\%$ confidence interval of $-0.241<\alpha<0.018$ ($\Delta\chi^2<1$ for one fitted parameter), indicating only a weak dependence of the reconnection rate on $Pr_{\rm M}$. In particular, a $Pr_{\rm M}$-independent reconnection rate ($\alpha=0$) is fully consistent with the measurements, whereas the commonly assumed scalings $\alpha=-1/4$ and especially $\alpha=-1/2$ provide progressively poorer fits. The latter yields $\Delta\chi^2=5.68$, making it significantly less consistent with the data than the nearly $Pr_{\rm M}$-independent scaling favoured by the simulations. This statistical analysis quantitatively supports the conclusion drawn from figure~\ref{fig:S5e5_PmDep} that, once the current sheet enters the fully plasmoid-mediated regime, the reconnection rate depends at most only weakly on the magnetic Prandtl number.

Having characterised the behaviour in our setup, we now turn to understanding why the resulting $Pr_\mathrm{M}$ dependence differs from that reported in studies of the Taylor problem. To this end, we perform a set of simulations using a configuration closely following CGW15.

\section{Comparison to the Taylor problem}
\label{sec:taylor}

{We now compare our results with the driven Taylor problem studied by \citet{Comisso2015} (hereafter CGW15). Their analysis predicts a strong magnetic Prandtl number dependence,
$V_{\rm rec}\propto (1 + Pr_{\rm M})^{-1/2}$ for $Pr_{\rm M}\gg1$ in the plasmoid-mediated regime. This is a steeper dependence
than the scaling derived by \citet{Park1984} for the Sweet-Parker regime.

CGW15 analysed the plasmoid-mediated regime of the forced Taylor problem. In their model, the global reconnection layer fragments into a hierarchy of secondary current sheets that remain close to marginal stability to the plasmoid instability. The dimensions of these local sheets are therefore not determined by the global Sweet-Parker balance, but by the condition that their effective Lundquist number remains close to the viscosity-dependent critical value for plasmoid onset. Combining this marginal-stability condition with the visco-resistive Sweet-Parker scaling of \citet{Park1984} for each secondary sheet yields the stronger dependence
$V_{\rm rec}\propto Pr_{\rm M}^{-1/2}$ for $Pr_{\rm M}\gg1$. Thus, the difference between the $Pr_{\rm M}^{-1/4}$ and $Pr_{\rm M}^{-1/2}$ scalings reflects the different current-sheet structures being assumed: a single steady laminar sheet in the former case and a hierarchy of marginal secondary sheets in the latter.}

Further details of the Taylor problem can be found in \citet{Comisso2015}. Here we summarise only those aspects of the setup that are relevant for comparison with the present work.

The equilibrium magnetic field is given by $B_y=x$, with perfectly conducting boundaries at $x=\pm L_x/2=\pm1$. This equilibrium is stable to the tearing instability, and reconnection is instead driven externally through a boundary perturbation of the form
\begin{equation}
x_w = \pm1 \mp \Xi(t)\cos(ky),
\end{equation}
where {$\Xi(t) = \Xi_0(1-e^{-t/\tau} - (t/\tau)e^{-t/\tau})$} 
with $\Xi_0=0.04$ and $\tau=2$ {and $x_w$ 
represents the instantaneous $x$-coordinate of the displaced conducting wall}.
Following CGW15, we choose a domain with $L_y=16\pi$ and perturb the fundamental mode,
\begin{equation}
k = \frac{2\pi}{L_y} = \frac 1 8.
\end{equation}

We use the \textsc{Pencil Code} to perform only two simulations in the region of parameter space identified by CGW15 as leading to plasmoid-dominated reconnection. Our numerical grid has a uniform resolution with $\Delta x=L_x/2048 \approx 9.76\times10^{-4}$ and $\Delta y = L_y/8192 = 6.13\times10^{-3}$. Both simulations use the same resistivity, $\eta = 4\times10^{-7}$, while the viscosity is varied such that the magnetic Prandtl numbers are $Pr_\mathrm{M}=10$ and $40$.
The reconnection rate is measured as 
\begin{equation}
    R(t) = \dfrac{d}{dt} \left[\max_{y} A_z(0,y,t) - \min_{y} A_z(0,y,t)\right],
\end{equation}
following equation~10 of \cite{Comisso2014}.

\begin{figure}
    \centering
    \includegraphics[width=0.75\linewidth]{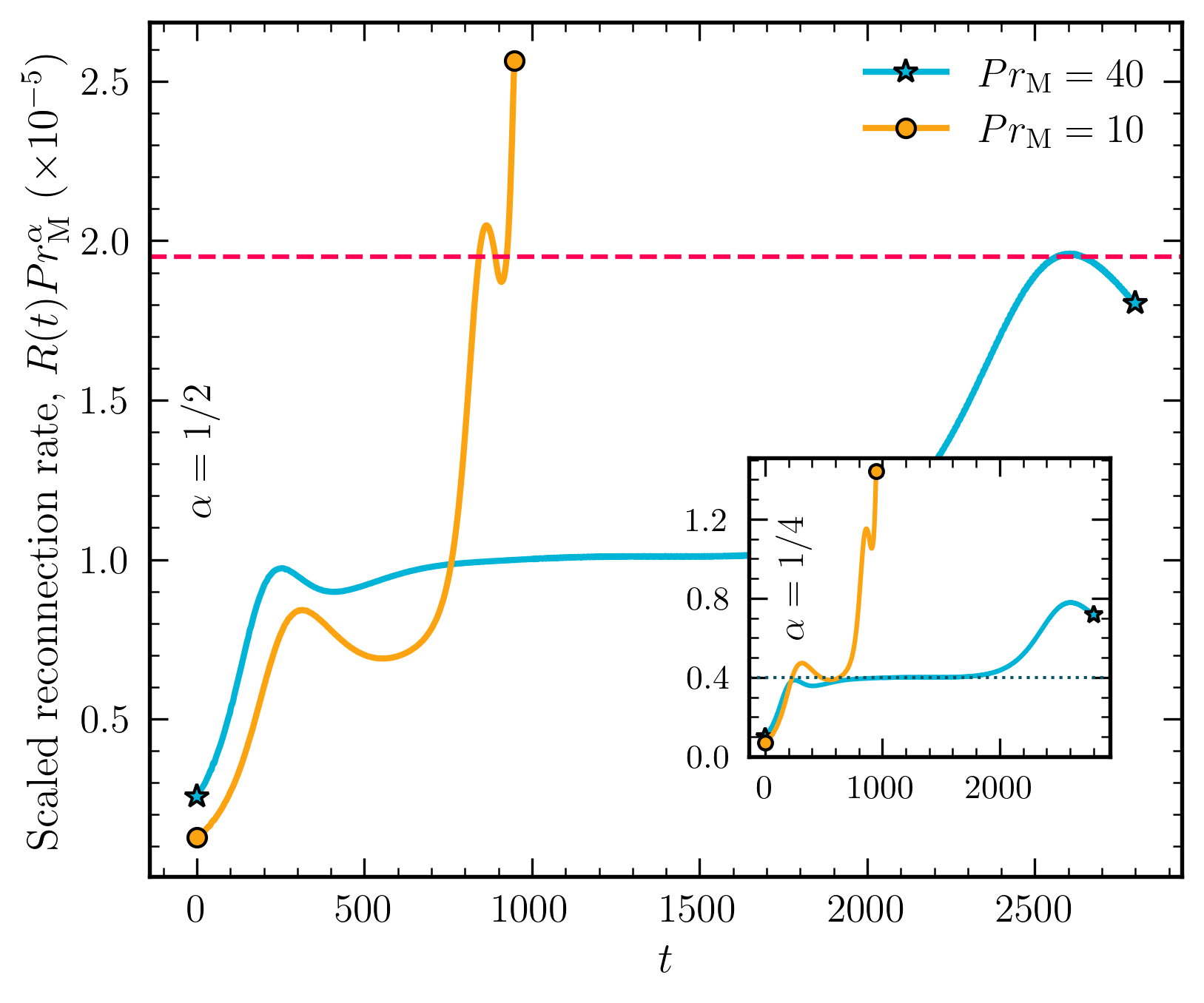}
    \caption{Time evolution of the reconnection rate for the two Taylor-problem simulations, following the setup of CGW15. In the main panel, the reconnection rate is rescaled by $Pr_\mathrm{M}^{1/2}$, while in the inset it is rescaled by $Pr_\mathrm{M}^{1/4}$. The former produces similar amplitudes in the plasmoid-dominated phase, whereas the latter gives a collapse during the Sweet-Parker-like phase. The horizontal reference lines (pink dashed in the main panel and black dotted in the inset) are shown only as visual guides.}
    \label{fig:driven-rate}
\end{figure}

Figure~\ref{fig:driven-rate} shows the reconnection-rate time series for the two driven reconnection simulations. Similar to CGW15, the system first develops an elongated Sweet-Parker-like current sheet, which subsequently becomes unstable to plasmoid formation. 

To examine the $Pr_\mathrm{M}$ dependence in the different stages of the evolution, we rescale the reconnection rate by different powers of $Pr_\mathrm{M}$. In the inset, where the reconnection rate is multiplied by $Pr_\mathrm{M}^{1/4}$, the curves show a reasonable collapse during the Sweet-Parker-like phase, consistent with the visco-resistive Sweet-Parker scaling obtained by \cite{Park1984} and CGW15. In contrast, during the plasmoid-dominated phase, a collapse of the reconnection rate is obtained when rescaled by $Pr_\mathrm{M}^{1/2}$, as shown in the main panel. This behaviour is again consistent with the prediction of CGW15 and \cite{Loureiro2013} in the plasmoid-mediated regime.

Both our simulations and those presented in CGW15 encounter numerical difficulties shortly after entering the plasmoid-mediated regime. This typically occurs once the plasmoids begin to coalesce, leading to the development of increasingly sharp gradients also in the $y$-direction. As a result, neither set of simulations is able to sustain a long statistically steady plasmoid-dominated state with ongoing plasmoid interactions and mergers.
We therefore identify two potentially important differences between our coalescing islands setup and the Taylor problem considered in CGW15.

First, in the Taylor problem, the equilibrium current sheet is stable to the tearing instability. Reconnection is initiated externally through boundary perturbations and therefore corresponds to a case of \emph{forced} reconnection. In contrast, the coalescing islands setup undergoes \emph{spontaneous} reconnection arising self-consistently from the non-linear evolution of the system.

Second, in CGW15 (as well as in our corresponding simulations reproducing their setup), the plasmoid-mediated reconnection rate is measured during the early plasmoid phase, before substantial mergers between individual plasmoids occur. By contrast, in our coalescing-islands setup, as discussed in subsection \ref{subsec:Pm-dep-plasmoid}, interactions and mergers between multiple plasmoids appear to be essential for attaining the large reconnection rates that become nearly independent of $Pr_{\rm M}$. This distinction may also help reconcile our results with the magnetic-Prandtl-number dependence of the plasmoid onset time shown in figure~7. Prior to the onset of the fast-reconnection phase, the current sheet is expected to undergo the initial tearing instability, producing plasmoids that remain largely isolated and evolve without significant mutual interaction. One possible interpretation is that the time required for these initially isolated plasmoids to grow to sizes at which they begin to interact is governed by the same visco-resistive physics that controls the early plasmoid-mediated reconnection rate measured by CGW15. If so, this would naturally account for the fact that the onset time follows the $Pr_{\rm M}$ scaling predicted by CGW15. Once the plasmoids have grown sufficiently to interact and merge, however, the dynamics enter a strongly nonlinear regime in which repeated plasmoid mergers substantially enhance the reconnection rate. Our results suggest that it is this merger-dominated regime that gives rise to the large reconnection rates exhibiting only a weak dependence on $Pr_{\rm M}$.

\section{Discussion}
\label{sec:discussion}

In this work, we have investigated the dependence of the reconnection rate on the magnetic Prandtl number in a two-dimensional setup consisting of two coalescing magnetic islands. Our primary motivation was to understand whether the reconnection scalings obtained in externally driven configurations, such as the Taylor problem, remain applicable in more self-consistent reconnecting structures resembling those observed in decaying MHD turbulence.

For Lundquist numbers below the onset of the plasmoid instability, our simulations recover the expected visco-resistive Sweet–Parker scaling, with the reconnection rate decreasing approximately as $Pr_\mathrm{M}^{-1/4}$. Once the current sheet becomes plasmoid unstable, however, the behaviour changes qualitatively. In the fully plasmoid-mediated regime ($S \gtrsim 10^5$), we find that the reconnection rate depends only weakly on $Pr_\mathrm{M}$ over the explored parameter range, in contrast to the $Pr_\mathrm{M}^{-1/2}$ scaling predicted for the plasmoid-mediated Taylor problem by CGW15.

A closer examination of the high-$S$ runs suggests that this difference may be connected to the dynamical state of the plasmoid population. In our coalescing islands setup, the largest reconnection rates are attained during phases in which multiple plasmoids interact, merge, and undergo continual non-linear evolution. By contrast, phases dominated by a single isolated plasmoid are associated with slower reconnection. This suggests that the reconnection dynamics in the strongly non-linear plasmoid-interaction regime may no longer be controlled by the same local Sweet–Parker balances that underlie the analytic $Pr_\mathrm{M}^{-1/2}$ prediction.

The comparison with the Taylor problem further supports this interpretation. In the boundary-driven setup, the equilibrium current sheet is initially stable and reconnection is triggered externally through imposed perturbations. Moreover, both our reproductions of the CGW15 setup and the original simulations themselves terminate shortly after the onset of plasmoid coalescence, owing to the development of increasingly sharp gradients. As a result, the reconnection rate is effectively measured during the early plasmoid phase, before a sustained non-linear state with strong plasmoid interactions is established. In contrast, the coalescing islands setup naturally evolves into a state with repeated plasmoid formation, interaction, and merger. The differing $Pr_\mathrm{M}$ scalings may therefore reflect a genuine physical distinction between these regimes, rather than a disagreement between numerical results.

Our results may also have implications for reconnection-mediated decay in magnetically dominated turbulence.
Recent studies of decaying MHD turbulence have reported decay timescales that appear to be largely independent of $Pr_\mathrm{M}$
at sufficiently high Lundquist numbers \citep{Brandenburg2024}.
If reconnection in such systems predominantly occurs through strongly non-linear plasmoid interactions similar to those observed here, then the weak $Pr_\mathrm{M}$ dependence found in our simulations may help explain these observations.

From a cosmological perspective, these results may also have implications for the evolution of primordial magnetic fields. For example, \cite{Hosking2023} estimated turbulent decay rates using the commonly assumed $Pr_\mathrm{M}^{-1/2}$ scaling of the reconnection rate when assessing the survivability of primordial magnetic fields in cosmic voids.
If the weak $Pr_\mathrm{M}$ dependence found here persists in fully turbulent systems, {and if magnetic reconnection is indeed the process governing the decay of primordial magnetic fields}, then such estimates may need to be revisited,
potentially modifying the resulting constraints on primordial magnetogenesis scenarios; see \cite{Neronov+24}. 
{Indeed, \citet{Brandenburg2024} found, in the context of decaying MHD turbulence, that the decay-time prefactor exhibits little or no additional dependence on the magnetic Prandtl number, although it retains a dependence on the Lundquist number over the range accessible to their three-dimensional simulations. Replacing the $Pr_\mathrm{M}^{1/2}$ enhancement of the decay time assumed by \citet{Hosking2023} with this approximately $Pr_\mathrm{M}$-independent behaviour, they estimated that, for $Pr_\mathrm{M}\sim10^7$, the magnetic field strength at recombination would shift from $\sim10^{-9}\,\mathrm{G}\,(\xi_M/1\,\mathrm{pc})$ to $\sim10^{-12.8}\,\mathrm{G}\,(\xi_M/1\,\mathrm{pc})$, where $\xi_M$ is the comoving magnetic integral scale. This substantially alters the inferred constraints on primordial magnetic fields, although the resulting field strengths remain above the lower limits inferred from blazar observations \citep{Neronov2010}. We emphasize, however, that our simulations probe only $1\le Pr_\mathrm{M}\le50$, and therefore any extrapolation of the weak $Pr_\mathrm{M}$ dependence found here to the cosmologically relevant regime of $Pr_\mathrm{M}\sim10^7$ should be regarded as tentative.}

Several important caveats remain. First, all simulations presented here are two-dimensional. In fully three-dimensional systems, reconnection may proceed through additional pathways involving kink-like distortions \citep{Oishi2015}, turbulence and stochastic field-line wandering \citep{Lazarian1999,Kowal2009,Kowal2012,Vicentin2025,Russell_2025}, and reconnection away from simple X-points. Recent work has shown that, at least at low Lundquist numbers, tearing-like instabilities can nevertheless persist in genuinely three-dimensional flux-tube-like magnetic configurations while still retaining several similarities with classical 2D tearing dynamics \citep{Kumar2025}. Determining whether the weak $Pr_\mathrm{M}$ dependence observed here survives in fully non-linear 3D plasmoid-mediated reconnection remains an important direction for future work.

Second, although our simulations reach Lundquist numbers well into the plasmoid-mediated regime, the explored parameter range remains limited compared to many astrophysical systems. Establishing the true asymptotic behaviour at simultaneously large $S$ and large $Pr_\mathrm{M}$ will require substantially higher-resolution simulations, likely in three dimensions.

Overall, our results suggest that the reconnection rate may become only weakly dependent on the magnetic Prandtl number once the plasmoid dynamics enter the strongly non-linear regime. Whether this behaviour persists at significantly larger Lundquist numbers and in fully three-dimensional turbulent systems remains an important question for future work.

\section*{Acknowledgements}
The simulations were carried out on the \emph{Dardel} supercomputing cluster under allocations granted by the Swedish National Infrastructure for Computing through the Center for Parallel Computers at the Royal Institute of Technology (KTH), Stockholm. 
We thank the reviewers for their careful reading of the manuscript and for their constructive suggestions, which have strengthened the manuscript and improved the clarity of its presentation. 
We also thank M.\ Rheinhardt (Aalto, Helsinki) for helpful comments on the manuscript. 
V.K.\ thanks P.\ Bhat (ICTS, Bengaluru), A.\ Bhattacharjee (Princeton), and A.\ Lazarian (Wisconsin–Madison) for insightful discussions. 
V.K.\ also thanks S.\ Kapoor (ICTS, Bengaluru), C.\ Rajarshi (ICTS, Bengaluru)  and Chandranathan\ A. (ICTS, Bengaluru) for constructive discussions during the revision process that helped shape several improvements incorporated into the revised manuscript.
V.K.\ gratefully acknowledges the hospitality and support of Nordita through the Nordita Visiting PhD Fellowship, as well as support from the Department of Atomic Energy, Government of India, under Project Nos.\ RTI4013 and RTI4019.
A.B.\ was supported in part by the European Research Council through the ERC Synergy Grant COSMOMAG under grant No.\ 101224803,
the Swedish Research Council (Vetenskapsr{\aa}det) under grant No.\ 2025-05957,
the National Science Foundation under grant Nos.\ NSF AST-2307698, AST-2408411, and NASA Award 80NSSC22K0825.

\section*{Data availability statement}

The data that support the findings of this study are openly available on
Zenodo at \href{https://doi.org/10.5281/zenodo.20268567}{doi:10.5281/zenodo.20268567} (v2026.05.18) or, for easier access, at
\url{http://norlx65.nordita.org/~brandenb/projects/plasmoid-mediated/}.
All calculations have been performed with the {\sc Pencil Code}
\citep{PC};
\href{https://doi.org/10.5281/zenodo.3961647}{doi:10.5281/zenodo.3961647}.

\begin{appendix}

\section{Pressure equilibrium vs complete equilibrium} \label{app:PressEq}

In \Sec{InitialConditions}, we mentioned an important subtlety regarding pressure equilibrium of the initial condition.
VKDL26 performed 2D simulations with a setup very similar to ours, using the same magnetic field profiles, in order to investigate the onset of fast, Lundquist-number-independent reconnection mediated by plasmoids. They reported the existence of an intermediate Lundquist number regime, $10^4 \lesssim S \lesssim 10^5$, in which the reconnection rate scales as $S^{-1/3}$. We do not observe such a regime in our simulations.

\begin{figure}
    \centering
    \includegraphics[width=0.75\linewidth]{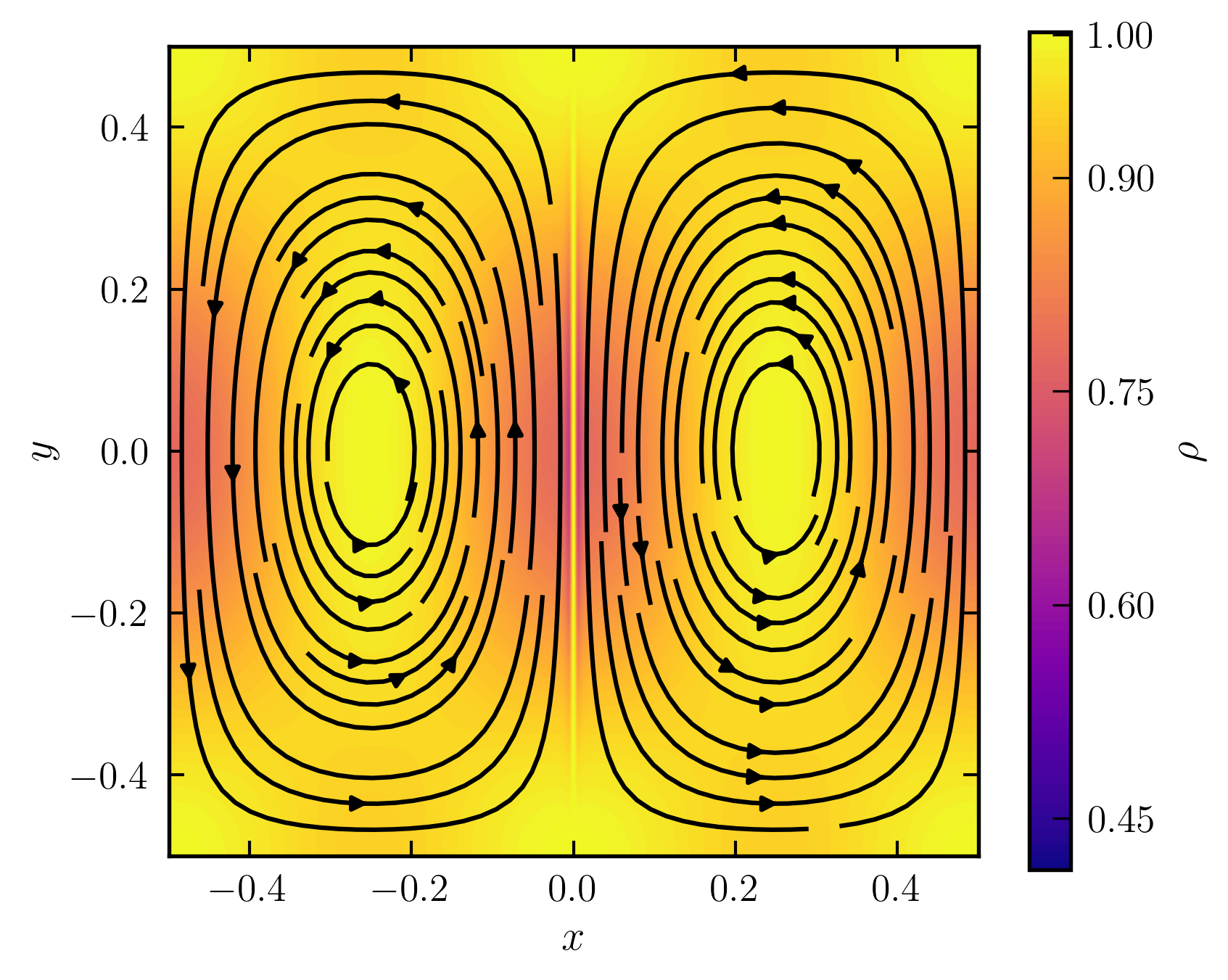}
    \caption{Density isocontours overlaid with magnetic field streamlines for the pressure equilibrium case. As in \Fig{fig:rho_B_stream}, this snapshot is taken from the run with $S=40000$ to enable a direct comparison.}
    \label{fig:rho_B_stream_peq}
\end{figure}

\begin{figure}
    \centering
    \includegraphics[width=1.0\linewidth]{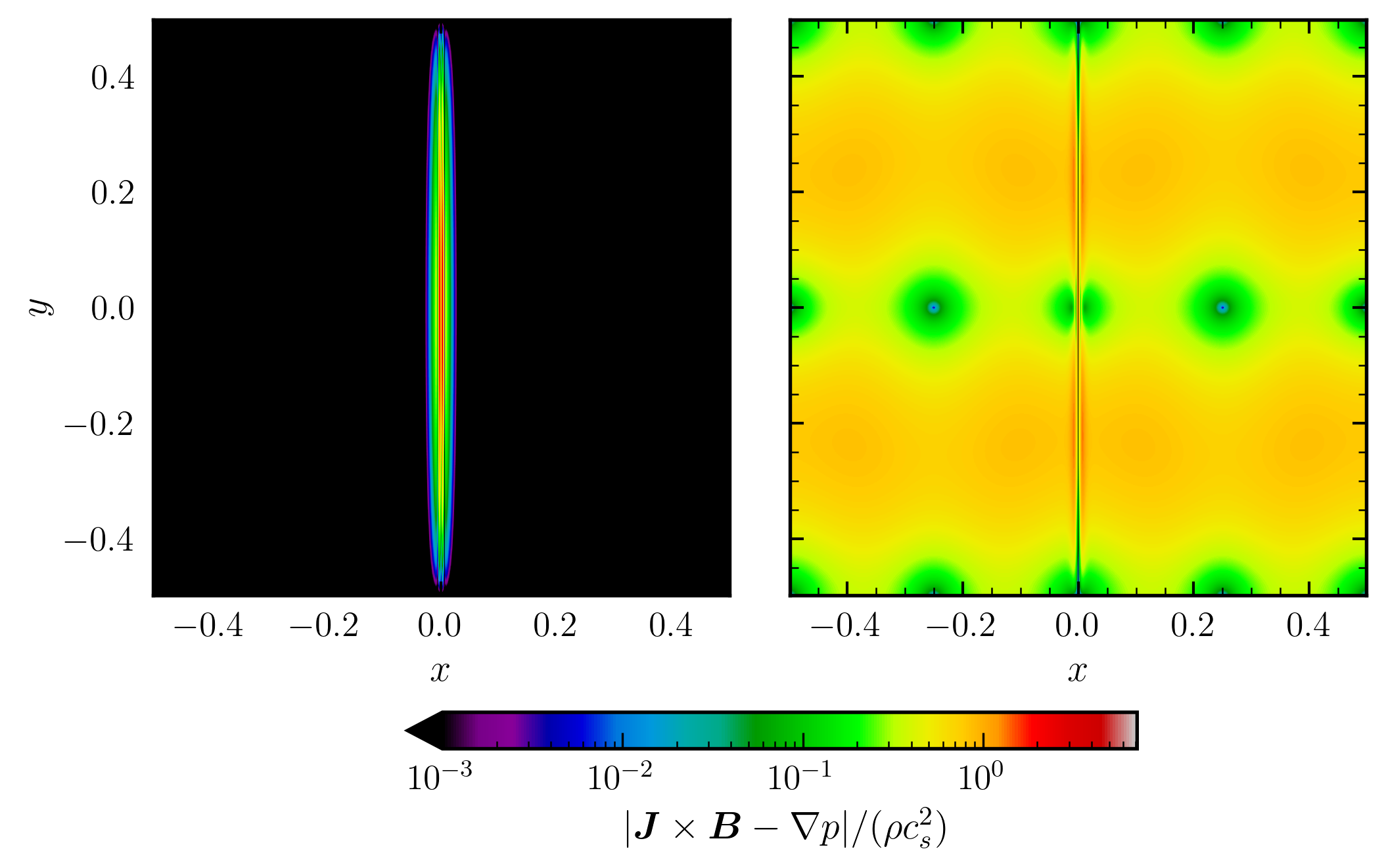}
    \caption{Comparison between the full equilibrium (left) and pressure-equilibrium-only (right) cases. The colours show the normalised net force,
    $|\bm{J}\times\bm{B}-\nabla p| / (\rho c_s^2)$, at $t=0$ for both cases.}
    \label{fig:force_init}
\end{figure}

{We argue that this discrepancy is most likely due} to two subtle but important differences between their setup and ours. In VKDL26, the initial density profile is chosen such that it balances only the magnetic pressure gradient. Figure~\ref{fig:rho_B_stream_peq} shows the corresponding density profile, which visually matches the profile used in VKDL26.
However, because the magnetic field lines are curved, the field also exerts magnetic tension forces. As a result, pressure balance alone does not correspond to a complete force equilibrium. The resulting force imbalance can become dynamically important in the low-$\beta$ limit. In VKDL26, $\beta \approx 2$, placing their simulations firmly in the low-$\beta$ regime where the residual force imbalance associated with magnetic tension can become dynamically important and potentially influence the observed reconnection-rate scaling. 
In contrast, our simulations have $\beta \approx 50$, for which these effects are expected to be much weaker.
We show in Figure~\ref{fig:force_init} the normalised magnitude of the difference between the total Lorentz force and the pressure gradient — a measure of the residual force imbalance — for both setups. The pressure-equilibrium-only configuration used in VKDL26 exhibits an order-unity imbalance over a substantial fraction of the domain, whereas in our approximately full-equilibrium setup this imbalance remains below $10^{-3}$ throughout almost the entire domain.

\begin{figure}
    \centering
    \includegraphics[width=0.75\linewidth]{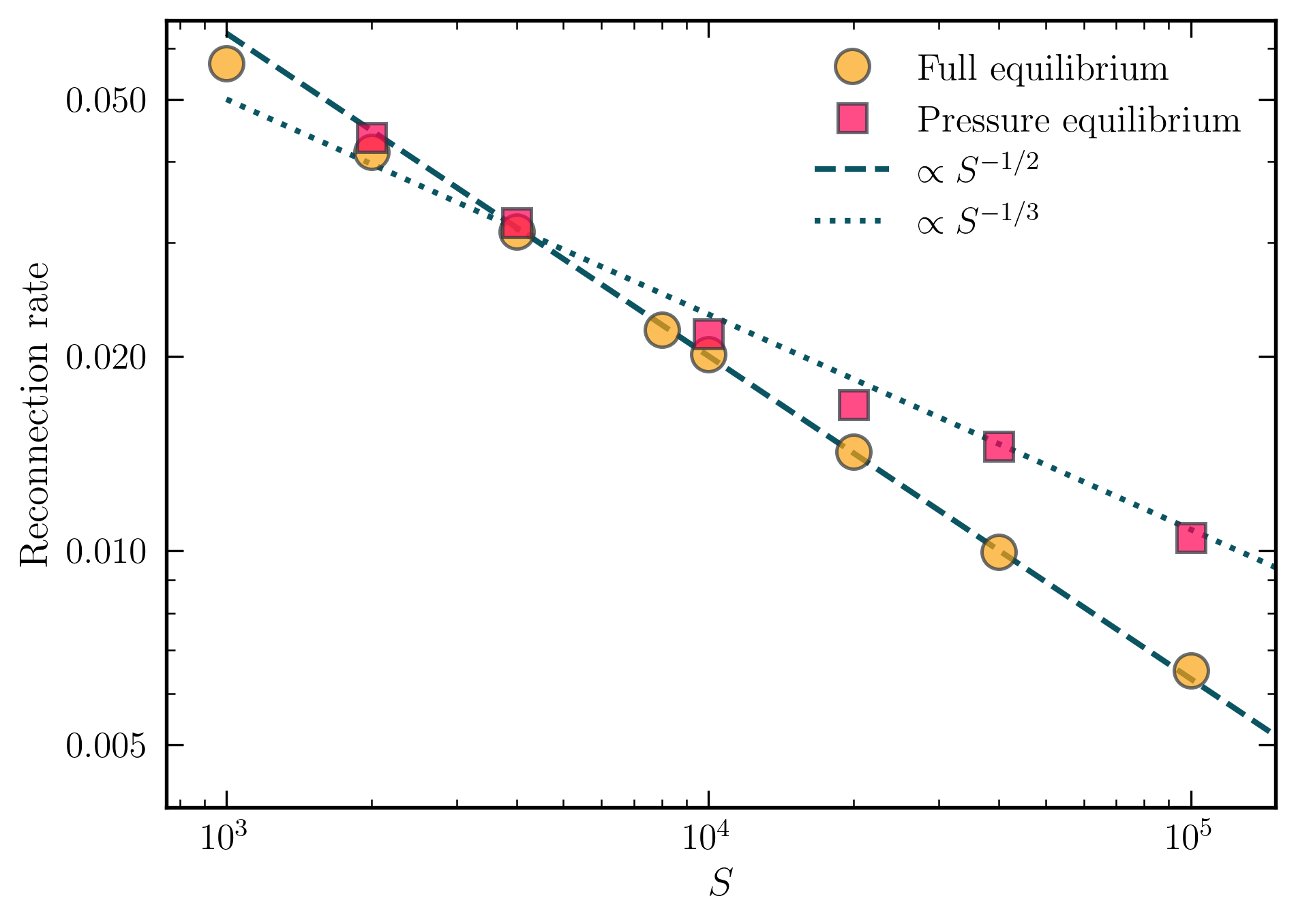}
    \caption{Comparison between reconnection rates as a function of Lundquist number with different initial density configurations at $Pr_{\rm M} = 1$. The configuration with full equilibrium shows a clean $S^{-1/2}$ scaling up to $S=10^5$, whereas the simulations with only pressure equilibrium show a different, $S^{-1/3}$ scaling for $S\gtrsim 10^4$, as obtained by VKDL26.}
    \label{fig:vrec_vs_S_peq}
\end{figure}

{To investigate whether this residual force imbalance could account for the discrepancy,} we also performed simulations using the pressure-equilibrium-only setup of VKDL26 (albeit without a guide field) with $\beta \approx 1.125$. This was achieved by setting the sound speed to $1.5$ in Alfv\'en units. We carried out simulations at $Pr_\mathrm{M}=1$ over a range of Lundquist numbers and extracted the corresponding reconnection rates. These results are shown in Figure~\ref{fig:vrec_vs_S_peq}, together with the results from the full-equilibrium setup (reproduced from { Figure~\ref{fig:reconnection_rate_scaling_S}}).

One can clearly see that the intermediate $S^{-1/3}$ scaling regime appears only in the pressure-equilibrium configuration, and in precisely the same Lundquist number range reported by VKDL26. {While this strongly suggests that the residual force imbalance associated with the pressure-equilibrium initial condition is responsible for the observed intermediate scaling, our comparison does not isolate this effect completely, as the pressure-equilibrium runs also differ from our standard setup in plasma $\beta$ and the absence of a guide field.}

Whether the pressure-equilibrium-only configuration or the approximately full-equilibrium configuration more accurately captures the conditions relevant to reconnecting structures in MHD turbulence and astrophysical plasmas is beyond the scope of the present work. Nonetheless, we expect our results in the fully plasmoid-mediated regime ($S \gtrsim 10^5$), where plasmoids strongly interact and merge with one another, to remain largely unaffected by the differences that give rise to the intermediate scaling regime over $10^4 \lesssim S \lesssim 10^5$.

\section{Measures of reconnection rate} \label{app:Rates}

In \Sec{Diagnostics}, we discussed different measures of the reconnection rate.
In this appendix, we explain our diagnostic for measuring it and compare with an alternative diagnostic used in the literature.
We also provide some physical intuition for the diagnostic adopted in this work.

We measure the rate of flux depletion at the current 
sheet as the reconnection rate. 
In two dimensions, with no out-of-plane magnetic field component, the magnetic field can be written in terms of the $z$-component of the vector potential as
\begin{equation}
B_x = \partial_y A_z, \qquad B_y = -\partial_x A_z,
\end{equation}
so that $A_z$ acts as a flux function. Since we impose perfectly conducting boundary conditions and initialise $A_z = 0$ at the domain boundaries, $A_z$ remains zero there for all times.

Using $\partial_x A_z = -B_y$, we write
\begin{equation}
A_z(x,y) = - \int_{-L_x/2}^{x} B_y(x',y)\, dx',
\label{eq:Az_intx}
\end{equation}
so that, along the current sheet ($x=0$), we have
\begin{equation}
A_z(0,y) = -\int_{-L_x/2}^{0} B_y(x',y)\, dx'.
\end{equation}

Using $\partial_y A_z = B_x$, we may equivalently write
\begin{equation}
A_z(x,y) = \int_{-L_y/2}^{y} B_x(x,y')\, dy',
\label{eq:Az_inty}
\end{equation}
and, across the current sheet ($y=0$), we have
\begin{equation}
A_z(x,0) = \int_{-L_y/2}^{0} B_x(x,y')\, dy'.
\end{equation}
The two expressions for $A_z(x,y)$, \Eqs{eq:Az_intx}{eq:Az_inty}
are consistent since $\nabla \cdot \BB = 0$. 
The magnetic flux contained within an island, $\Phi_{\mathrm{island}}$ is proportional to the difference between the extrema of $A_z(x,0)$ and $A_z(0,y)$. The instantaneous non-dimensional reconnection rate is therefore defined as
\begin{equation}
V_{\mathrm{rec}} = -\dfrac{1}{V_A\,B_\mathrm{amp}}\dfrac{d}{dt}
\left[
\max_{x} A_z(x,0) \;-\; \max_{y} A_z(0,y)
\right].
\end{equation}
This definition is identical to that used by 
\cite{Bhattacharjee2009} and HB10, except that those works assumed 
\begin{equation}
\partial_t\!\left[\max_{x} A_z(x,0)\right] \approx 0,
\end{equation}
which is well satisfied, particularly in high-$S$ runs. Under this approximation, the reconnection rate reduces to
\begin{equation}
V_{\mathrm{rec}} = \dfrac{1}{V_A\,B_\mathrm{amp}}\dfrac{d}{dt}
\left[
\max_{y} A_z(0,y)
\right].
\end{equation}

Our diagnostic is also closely related to that used by \cite{Vicentin2026} (VKDL26 hereafter), 
where the total unsigned magnetic flux is computed across a line 
perpendicular to the current sheet spanning the full domain 
(the $y=0$ line for an orientation similar to ours). 
This quantity is approximately equal to $4\Phi_{\mathrm{island}}$, 
whose time derivative, when appropriately normalised, gives the reconnection rate. 

\begin{figure}
    \centering
    \includegraphics[width=0.65\linewidth]{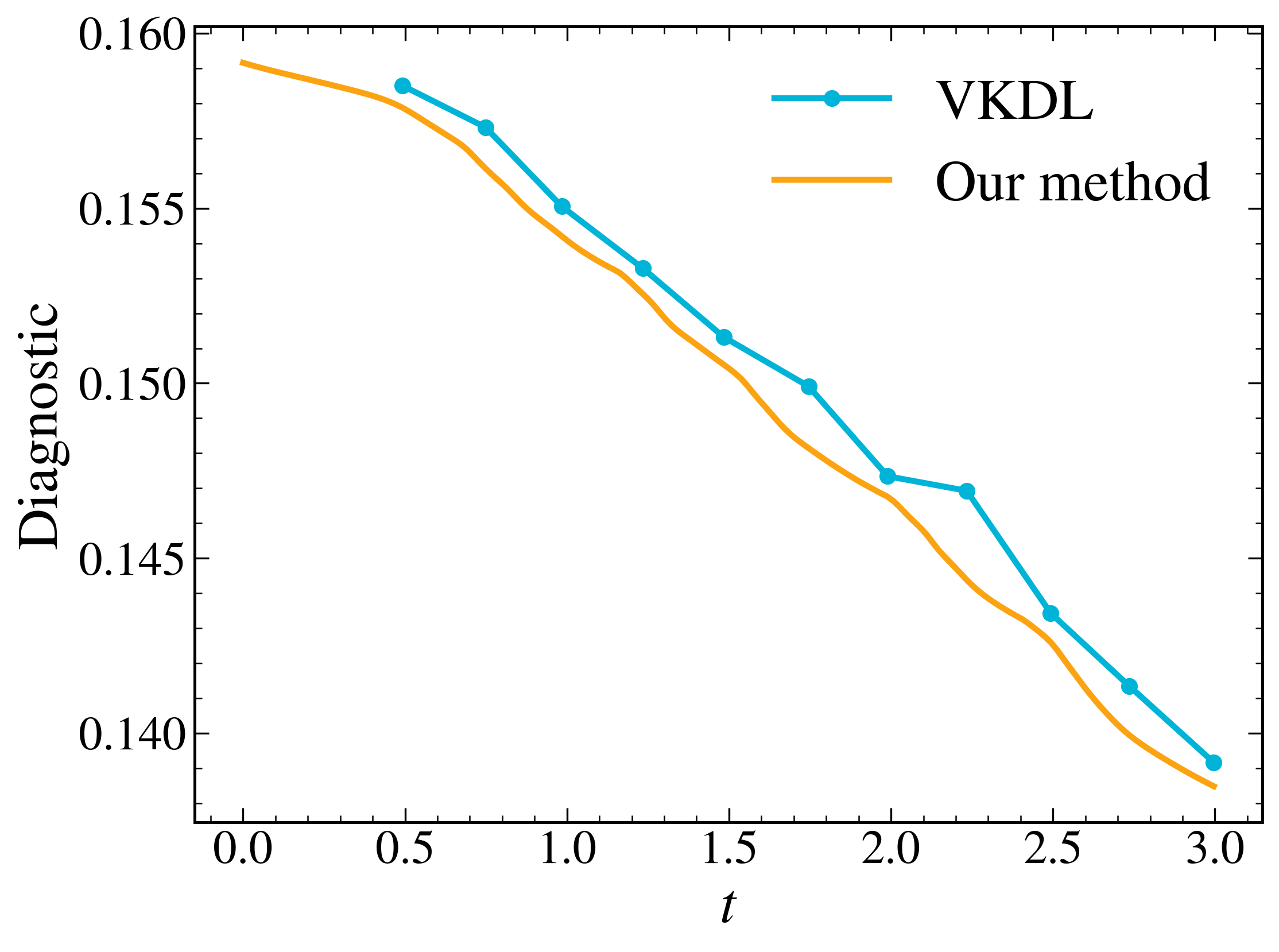}
    \caption{Comparison between the diagnostic quantity used by us and by VKDL26.}
    \label{fig:compare_diagnostic_16384h_2}
\end{figure}
Figure~\ref{fig:compare_diagnostic_16384h_2} compares the two flux measures as functions of time. Their appropriately normalised time derivatives yield the reconnection rate, and the close agreement between the two measures demonstrates the consistency of the diagnostics.

\section{Convergence of our simulations} \label{app:convergence}

\begin{figure}
    \centering
    \includegraphics[width=1\linewidth]{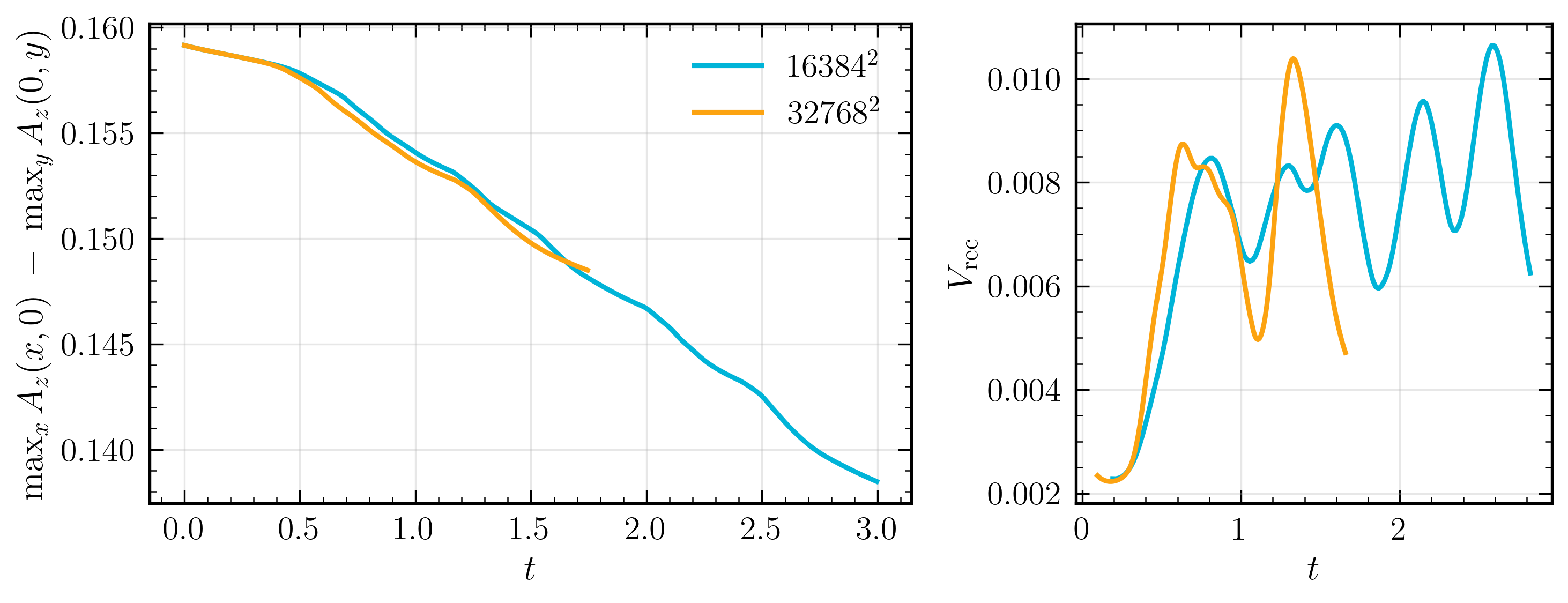}
    \caption{Flux evolution and corresponding reconnection rate for the runs at $S=5\times10^5$ and $Pr_{\rm M}=1$ at numerical resolutions of $16384^2$ and $32768^2$.}
    \label{fig:convergence}
\end{figure}

To verify that our measured reconnection rates are not influenced by numerical resolution, we repeated the $Pr_{\rm M}=1$ simulation at the highest Lundquist number considered in this work, $S=5\times10^5$, using resolutions of $16384^2$ and $32768^2$. Figure~\ref{fig:convergence} compares the evolution of the reconnected flux and the corresponding reconnection rate.
The two simulations exhibit very similar flux evolution and reconnection rates. The small differences that develop are attributed to the different realizations of the initial perturbations at the two resolutions. 

\begin{figure}
    \centering
    \includegraphics[width=1\linewidth]{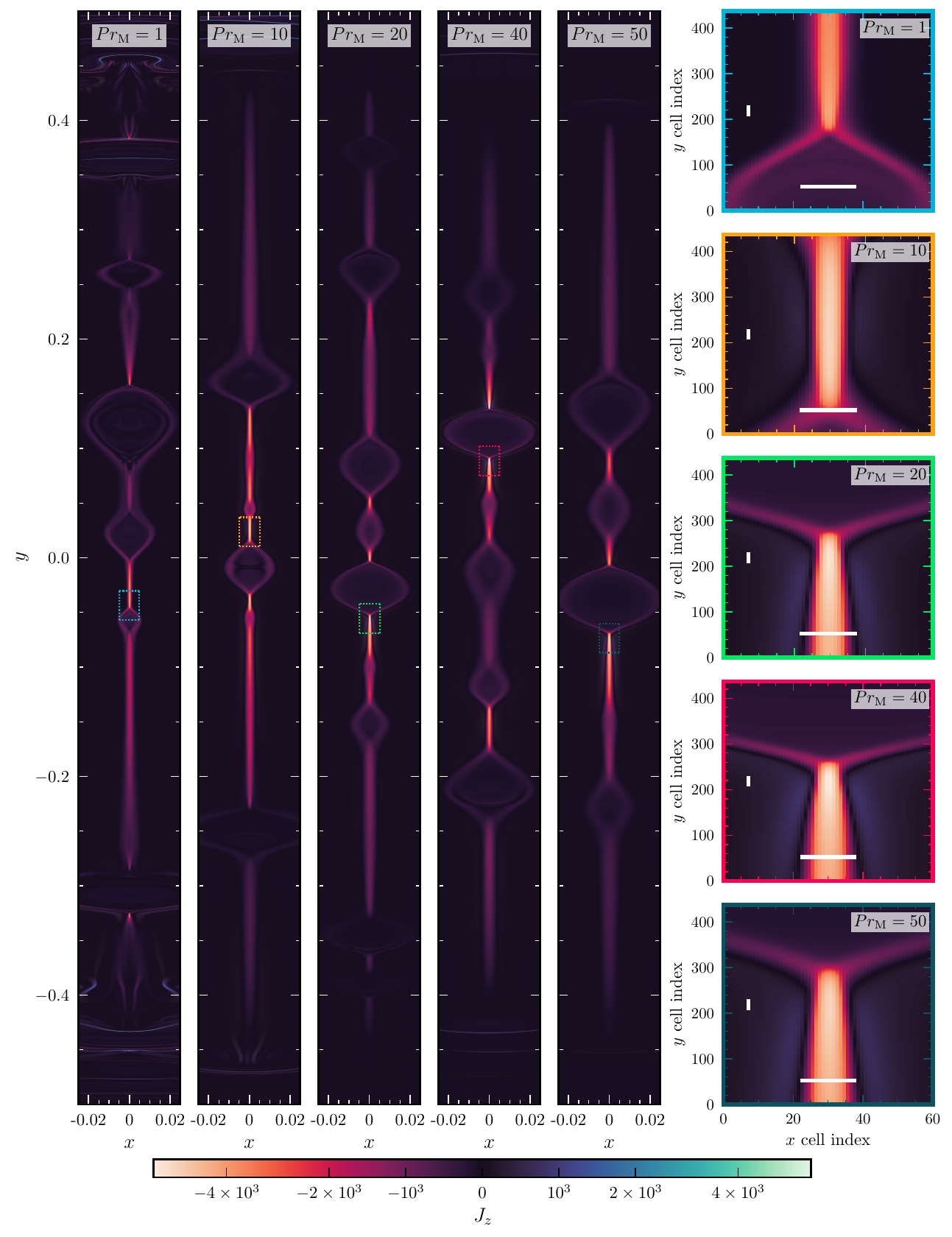}
    \caption{Current-density structure during plasmoid-mediated reconnection for different magnetic Prandtl numbers. Left: snapshots of the out-of-plane current density, $J_z$, for $Pr_{\rm M}=1$, $10$, $20$, $40$, and $50$, shown at representative times during the non-linear plasmoid regime. The dashed rectangles indicate the regions enlarged in the right-hand panels, each centred on the location of the maximum current density magnitude, $|J_z|$. The zoomed views demonstrate that the current sheets remain well resolved across the entire range of magnetic Prandtl numbers considered. The white vertical and horizontal lines in each right-hand panel correspond to a length of 15 grid cells.}
    \label{fig:plasmoids_pm}
\end{figure}

Owing to the substantially increased computational cost, we did not perform equivalent convergence tests for the higher-$Pr_{\rm M}$ simulations. However, increasing $Pr_{\rm M}$ is not expected to introduce finer spatial structures than in the $Pr_{\rm M}=1$ case. The higher-$Pr_{\rm M}$ runs should therefore be at least as well resolved as the converged $Pr_{\rm M}=1$ simulation. This expectation is further supported by \Fig{fig:plasmoids_pm}, where the zoomed views show that the peak-current region is resolved by sufficiently many grid cells for all values of $Pr_{\rm M}$ considered, with no evidence of Gibbs ringing or other grid-scale numerical artefacts.

We therefore conclude that the reconnection rates reported in this work are numerically converged and are not significantly affected by the adopted grid resolution.

\section{Three-parameter fits to the plasmoid-mediated reconnection rate} \label{app:fit_models}

In addition to the two-parameter model adopted in \Sec{subsec:Pm-dep-plasmoid},
\begin{equation}
V_{\rm rec}=V_0(1+Pr_{\rm M})^\alpha,
\end{equation}
we also explored the more general three-parameter form
\begin{equation}
V_{\rm rec}
=
V_0
\left(
{Pr_{\rm M0}+Pr_{\rm M}}
\right)^\alpha,
\end{equation}
where $V_0$, $\alpha$, and $Pr_{\rm M0}$ were all treated as free parameters.

As it turns out, the three-parameter fit does not provide a meaningful constraint on
$Pr_{\rm M0}$. Regardless of the initial guess, the optimisation consistently
converges to the largest value of $Pr_{\rm M0}$ permitted by the fitting
bounds, indicating that the available data do not contain sufficient
information to determine this additional parameter. The preference for large
values of $Pr_{\rm M0}$ is also consistent with the weak
magnetic-Prandtl-number dependence inferred in the main text: for
$Pr_{\rm M0}\gg Pr_{\rm M}$, the model
$V_{\rm rec}=V_0(Pr_{\rm M0}+Pr_{\rm M})^\alpha$ approaches an approximately
$Pr_{\rm M}$-independent form over the range of $Pr_{\rm M}$ considered here.
The present data therefore do not justify the introduction of the additional
parameter $Pr_{\rm M0}$.

\begin{table}
\centering
\caption{Best-fit parameters for the model
$V_{\rm rec}=V_0(Pr_{\rm M0}+Pr_{\rm M})^\alpha$
with fixed values of $Pr_{\rm M0}=0$, $1$, and $50$.
The quoted limits correspond to the lower and upper bounds of the
68\% ($1\sigma$) confidence interval for $\alpha$.}
\label{tab:fit_models}
\begin{tabular}{lcccc}
\hline
Model & $V_0$ & $\alpha$ & $\alpha_{\rm min}$ & $\alpha_{\rm max}$ \\
\hline
$V_0Pr_{\rm M}^{\alpha}$ &
$7.833\times10^{-3}$ &
$-0.085$ &
$-0.193$ &
$0.017$ \\

$V_0(1+Pr_{\rm M})^{\alpha}$ &
$8.458\times10^{-3}$ &
$-0.107$ &
$-0.241$ &
$0.018$ \\

$V_0(50+Pr_{\rm M})^{\alpha}$ &
$1.016\times10^{-1}$ &
$-0.649$ &
$-1.416$ &
$0.021$ \\
\hline
\end{tabular}
\end{table}

To illustrate this point, we also performed fits with $Pr_{\rm M0}$ fixed to a
set of representative values. Table~\ref{tab:fit_models} lists the resulting
best-fit values of $V_0$ and $\alpha$, together with the corresponding
$68\%$ confidence interval for $\alpha$, for $Pr_{\rm M0}=0$, $1$, and $50$.
The fits with $Pr_{\rm M0}=0$ and $1$ yield very similar values of the
scaling exponent, while adopting the much larger value
$Pr_{\rm M0}=50$ produces a steeper best-fit exponent but with a substantially
broader confidence interval that still encompasses a weak or negligible
$Pr_{\rm M}$ dependence. In fact, fixing $Pr_{\rm M0}=50$ requires a best-fit exponent of
$\alpha=-0.649$, which is even steeper than the
$\alpha=-1/2$ scaling predicted by \citet{Comisso2015} and substantially
steeper than the $\alpha=-1/4$ scaling of \citet{Park1984}. The increasingly steep best-fit exponent obtained as larger values of $Pr_{\rm M0}$ are imposed illustrates the strong degeneracy between $Pr_{\rm M0}$ and $\alpha$, indicating that the available data do not provide sufficient leverage to determine a characteristic $Pr_{\rm M0}$ independently.

\end{appendix}

\section*{Authors' ORCIDs}

\noindent
V. Kumar, https://orcid.org/0009-0008-2158-3774

\noindent
A. Brandenburg, https://orcid.org/0000-0002-7304-021X

\bibliography{pmref}{}
\bibliographystyle{jpp}
\end{document}